\documentclass[preprint,superscriptaddress,floatfix,preprintnumbers,nofootinbib,altaffilletter]{revtex4-1}
\pdfoutput=1
\usepackage[utf8]{inputenc}
\usepackage[colorlinks=true,citecolor=red,linkcolor=blue]{hyperref}
\usepackage{amsmath,amssymb}
\usepackage{epsfig}
\usepackage{graphicx}                          
\usepackage{url}
\usepackage{color}
\usepackage{slashed}
\usepackage{multirow}
\usepackage{placeins}
\usepackage[dvipsnames]{xcolor}
\usepackage{epstopdf}
\usepackage{mhchem}
\usepackage{blindtext}

\begin{document}

\title{A light complex scalar for the electron and muon anomalous magnetic moments }

\author{Jia Liu}
\affiliation{\mbox{Physics Department and Enrico Fermi Institute, University of Chicago, Chicago, IL 60637}}

\author{Carlos E.M. Wagner}
\affiliation{\mbox{Physics Department and Enrico Fermi Institute, University of Chicago, Chicago, IL 60637}}
\affiliation{\mbox{High Energy Physics Division, Argonne National Laboratory, Argonne, IL 60439}}
\affiliation{\mbox{Kavli Institute for Cosmological Physics, University of Chicago, Chicago, IL 60637}}

\author{Xiao-Ping Wang}
\affiliation{\mbox{High Energy Physics Division, Argonne National Laboratory, Argonne, IL 60439}}

\date{\today}
\preprint{EFI-18-16}

\begin{abstract}

The anomalous magnetic moments of the electron and the muon are interesting observables, since
they can be measured with great precision and their values can be computed with excellent accuracy
within the Standard Model (SM). The current experimental measurement of this quantities show a deviation
of a few standard deviations with respect to the SM prediction, which may be a hint of new physics. 
The fact that the electron and the muon masses differ by two orders of magnitude and the deviations have 
opposite signs makes it difficult to find a common origin of these anomalies. In this work we introduce a 
complex singlet scalar charged under a Peccei–Quinn-like (PQ) global symmetry together with the 
electron transforming chirally under the same symmetry. In this realization,  the CP-odd scalar 
couples to electron only,  
while the CP-even part can couple to muons and electrons simultaneously. In addition, the CP-odd scalar can 
naturally be much lighter than the CP-even scalar, as a pseudo-Goldstone boson of the PQ-like symmetry, 
leading to an explanation of the suppression of the electron anomalous magnetic moment with respect to the 
SM prediction due to the CP-odd Higgs effect dominance, as well as an enhancement of the muon one induced 
by the  CP-even component.    

\end{abstract}

\maketitle

\tableofcontents

\section{Introduction}

The Standard Model (SM) provides a precise theoretical framework for the description of all known 
interactions in nature.  The SM description of the interaction of quarks and leptons with electroweak
gauge  bosons has been probed at the per-mille level, being hence sensitive to quantum 
corrections to the tree-level results~\cite{Tanabashi:2018oca}. No significant deviations from the SM 
predictions have been found. 

Since Schwinger's first computation of the electron anomalous magnetic moment of the electron, it was realized
that its measurement can provide an accurate test of Quantum Electrodynamics (QED), and subsequently of the SM, 
describing the interactions of fundamental  particles in nature. The QED contribution \cite{Schwinger:1948iu,Sommerfield:1957zz,Petermann:1957hs,PhysRevLett.47.1573,Kinoshita:1990wp, Laporta:1996mq, Degrassi:1998es, Kinoshita:2004wi,Kinoshita:2005sm,Passera:2006gc,Kataev:2006yh,Passera:2006gc, Aoyama:2007mn,
Aoyama:2012wj, Aoyama:2012wk, Laporta:2017okg, Aoyama:2017uqe, Volkov:2017xaq, Volkov:2018jhy} to the anomalous magnetic moment of the 
electron and the muon is today known up to 5-loop order \cite{Tanabashi:2018oca,Mohr:2000ie,Czarnecki:1998nd}. 

The QED contribution, although dominant, is not the only one affecting the anomalous magnetic moments. 
The hadronic contributions \cite{Jegerlehner:1985gq, Lynn:1985sq, Swartz:1994qz, Martin:1994we, Eidelman:1995ny, Krause:1996rf, Davier:1998si, Jegerlehner:1999hg, Jegerlehner:2003qp, Melnikov:2003xd,
deTroconiz:2004yzs, Bijnens:2007pz, Davier:2007ua} become quite relevant and can be accurately computed from dispersion 
relations describing the electron-positron collisions with hadrons in the final states.  Moreover, the weak
interaction effects \cite{Czarnecki:1995wq, Czarnecki:1995sz, Czarnecki:1996if, Czarnecki:2002nt, Heinemeyer:2004yq, Gribouk:2005ee}, although suppressed by powers of the weak gauge boson masses,  become also relevant
at the level of accuracy provided by today's computations. Finally, there is a component of the hadronic
contribution, the so-called light-by-light contribution \cite{Bijnens:1995xf,Hayakawa:1997rq,Knecht:2001qf,Knecht:2001qg,RamseyMusolf:2002cy,Melnikov:2003xd,
	deTroconiz:2004yzs,Prades:2009tw, Kataev:2012kn, Kurz:2015bia, Colangelo:2017qdm}, 
which cannot be obtained experimentally and hence has to be estimated by theoretical methods.

Quite importance for these determinations is an accurate measurement of the fine structure constant. 
The authors of Ref.~\cite{Parker191} use the recoil frequency of
Cesium-133 atoms in a matter-wave interferometer to determine the mass of the Cs atom, and
obtain the most accurate  value of the fine structure constant to date.
By combining it with theory \cite{Aoyama:2014sxa, Mohr:2015ccw}, they obtain the electron magnetic dipole 
moment to be
\begin{align}
\Delta a_e \equiv a_e^{\rm exp} - a_e^{\rm SM} = (-88 \pm 36)\times 10^{-14},
\label{eq:g-2-e}
\end{align}
which implies the deviation has a negative sign and presents a $2.4~\sigma$ discrepancy 
\cite{Parker191, Jegerlehner:2018zrj, Davoudiasl:2018fbb} between the 
SM prediction and experimental measurements \cite{PhysRevLett.100.120801, Hanneke:2010au}.
On the other hand, the muon magnetic dipole moment has $3.7~\sigma$ discrepancy with a positive sign, 
opposite to the $a_e$ deviation~\cite{Blum:2018mom,Bennett:2006fi},
\begin{align}
\Delta a_\mu \equiv a_\mu^{\rm exp} - a_\mu^{\rm SM} = (2.74 \pm 0.73)\times 10^{-9}.
\label{eq:g-2-mu}
\end{align}

The $a_\mu$ deviation is of the same order of the weak corrections and hence can be naturally explained
by physics at the weak scale.  As it was first stressed in Ref.~\cite{Giudice:2012ms},
assuming similar corrections to $a_e$, due to the dependence on the
square of lepton mass, they become of the order of $\Delta a_e \simeq 0.7 \times 10^{-13}$. Therefore, they cannot  lead to an explanation of the $a_e$ anomaly. 
Moreover, if the interactions affecting electron and muon sector would be the same, one would
expect deviations of the same sign and not of opposite signs as observed experimentally, 
Eqs.~(\ref{eq:g-2-e}) and (\ref{eq:g-2-mu}).

To simultaneously explain the two anomalies, the interactions should violate lepton flavor universality in a 
delicate way, to contribute negatively for electrons while positively for muons. Recently,  the authors of 
Ref.~\cite{Davoudiasl:2018fbb} have provided a solution with one CP-even real scalar coupled to both $e$ and
$\mu$ with different couplings. To achieve negative contribution to $g-2$ of electron, they further
require that this scalar  contribute to $a_e$ via a 2-loop Barr-Zee diagram with the sign of the coupling 
specifically chosen to lead to the require effect.  Another recent work~\cite{Abu-Ajamieh:2018ciu}, 
also discusses both scalar and pseudo-scalar with 2-loop Barr-Zee, Light-By-Light and Vacuum Polarization
diagrams. In an independent work, the authors of Ref.~\cite{Crivellin:2018qmi} have, instead, added
both $SU(2)_{L}$ doublet and singlet vector-like heavy leptons,  which couple to the SM leptons via Yukawa interaction. 
The origin of different sign to $\Delta a_{e/\mu}$ comes from the sign of the off-diagonal Yukawa
coupling between heavy lepton and SM lepton.  

In this work, we shall assume that the reason for the discrepancy in sign of the deviations of $a_e$ and
$a_\mu$ with respect to the SM has to do, in part, with a difference in mass of the bosons interacting with 
these  particles at the loop level.  Moreover, we shall assume these bosons to proceed from a singlet
complex scalar, with electrons coupling to the CP-odd and CP-even components in a similar way,
but with the CP-odd effects becoming dominant due to the small mass of the corresponding scalar. On
the other hand, we shall assume that the muons interact mainly with the CP-even component.  We
shall achieve these properties by imposing an appropriate PQ-like symmetry, under which both the
complex scalar and the electron are charged. The CP-odd component may be
hence naturally light, since it could be a pseudo-Goldstone boson of the PQ-like symmetry.
The explanation  of the deviation of $a_\mu$, on the other hand, is similar to the one proposed
in several works in the literature~\cite{Kinoshita:1990aj, Zhou:2001ew, Barger:2010aj, TuckerSmith:2010ra, 
Chen:2015vqy, Liu:2016qwd, Batell:2016ove, Marciano:2016yhf, Wang:2016ggf}.

This article is organized as follows. In section~\ref{sec:g-2}, we describe the scalar and
pseudo-scalar corrections to the anomalous magnetic moments of the electron and muon. 
In section~\ref{sec:EFTModel}, we present an effective field theory description of our model,
describing the interactions of the leptons with the complex scalar after imposing the
PQ-like symmetry. In section~\ref{sec:UVModel}, we present an ultraviolet (UV) completion of the 
effective theory. In section~\ref{sec:discussionElecMuon}, we discuss the phenomenology
constraints on the UV complete model.
We reserve section~\ref{sec:conclusions} for our conclusions.

\section{g-2 anomalies for electron and muon}
\label{sec:g-2}

In our approach, the new physics only comes from the scalar sector, where a singlet light complex scalar 
$\phi$ solves both $\Delta a_{e/\mu}$. We use the fact that the contributions to $g-2$ of scalars with scalar and 
pseudo-scalar coupling to leptons  are of opposite sign. The pseudo-scalar $\phi_I$ from $\phi$ contributes
only to $\Delta a_e$  because of a global PQ-like symmetry and the CP symmetry, while the CP-even scalar 
$\phi_R$ is responsible for the contributions to $\Delta a_\mu$.  Therefore, the relative sign between $\Delta a_e$
and $\Delta a_\mu$ has its origin from the CP properties of scalars.

In the following we begin with a generic Yukawa coupling of a scalar to electron or muon. 
To be specific,  a scalar  with  both scalar and pseudo-scalar couplings to leptons,
$ S \bar{\ell} \left(g_R + i g_I \gamma_5 \right) \ell $,
it can contribute to the anomalous magnetic dipole moment as \cite{PhysRevD.5.2396, Leveille:1977rc}
\begin{align}
\Delta a_\ell = \frac{1}{8 \pi^2}   \int^1_0 dx 
\frac{(1-x)^2  \left((1+x)g_R^2 - (1-x) g_I^2 \right)}{(1-x)^2+x \left( m_S/m_\ell \right)^2} .
\end{align}

However, if a real scalar has both non-zero scalar and pseudo-scalar couplings, $g_R$ and $g_I$, respectively, 
the CP is violated and lepton electric dipole moment will be generated. To avoid this constraint, 
we  require CP conservation that each scalar has either scalar or pseudo-scalar couplings. 
In particular, we assume the presence of a
pseudo-scalar $\phi_I$ that couples to electron and a CP-even scalar which couples to muon as
\begin{align}
\mathcal{L}_{\rm int} = i g_{\phi_I}^e  \phi_I \bar{e} \gamma_5 e + g^\mu_{\phi_R} \phi_R \bar{\mu} \mu .
\end{align}

\begin{figure}
	  \begin{tabular}{cc}
	\includegraphics[width=0.46 \columnwidth]{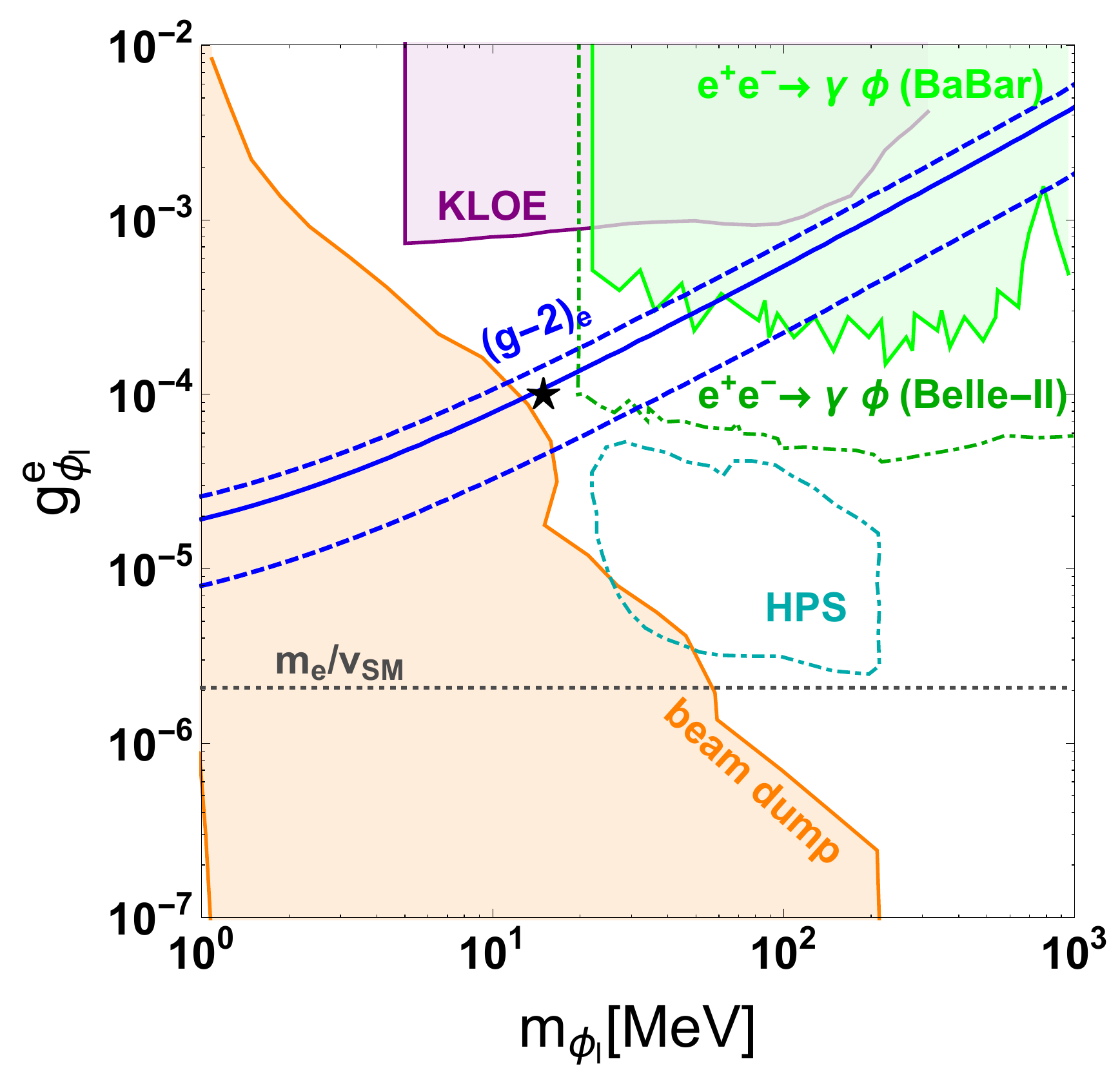} &
	\includegraphics[width=0.46 \columnwidth]{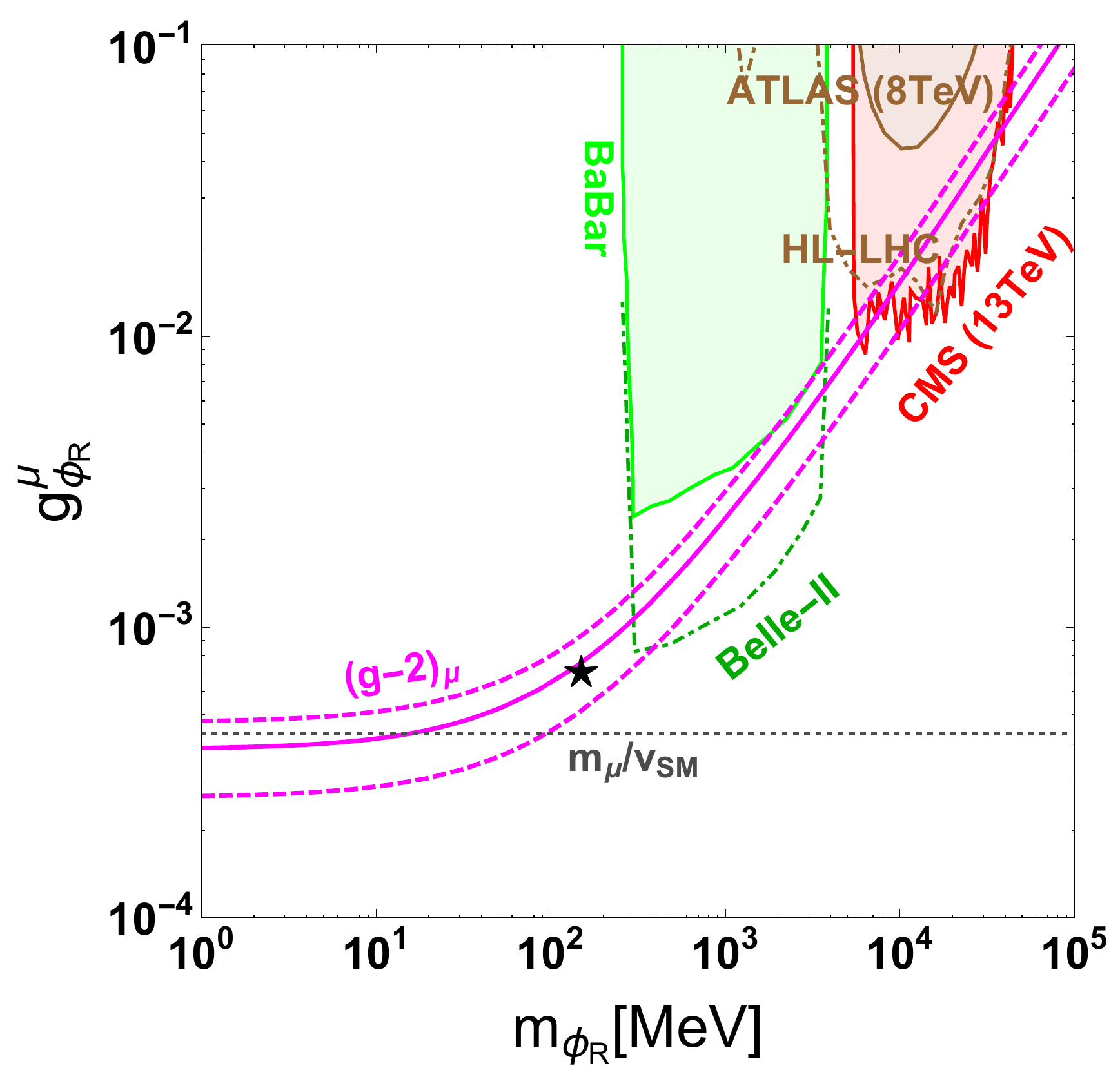} \\
	    (a) & (b)
\end{tabular}
	\caption{The color shaded regions with solid boundary are excluded by current experiments, the regions with
	dot-dashed boundaries are future projections. The black star corresponds to the benchmark
	in Table.~\ref{tab:benchmark1}. 
	(a): The parameter space $(g_{\phi_I}^e,  m_{\phi_I} )$ for $\Delta a_{e}$ and the constraints from different experiments.
	The shaded orange region is from beam dump experiment~\cite{Batell:2016ove, Liu:2016qwd} and the dot-dashed dark cyan contour area is from future projection for HPS \cite{Battaglieri:2014hga, Batell:2016ove}. The collider limits include shaded green region searching for $e^+ e^- \to \gamma \phi$ at BaBar  \cite{Knapen:2017xzo}, 
	shaded purple region from KLOE \cite{Anastasi:2015qla,Alves:2017avw} and Belle-II projection \cite{Abe:2010gxa, Batell:2016ove} which is shown in dot-dashed green contour region. (b): The parameter space $(g_{\phi_R}^\mu, ~ m_{\phi_R} )$ for $\Delta a_{\mu}$ and the 
	constraints from collider searches. BaBar search via $e^+ e^- \to \mu^+ \mu^- \phi$ is shown in the shaded green region
	\cite{Batell:2016ove, Batell:2017kty} and future projection for Belle-II \cite{Batell:2017kty} is shown by the
	green dot-dashed contour. 
    The ATLAS experiment has looked for exotic Z decay $Z \to 4\mu$ at LHC Run-I, which
    has been re-casted for scalar mediator by Ref.~\cite{Batell:2017kty} , and the limits for both Run-I 
    and HL-LHC are shown by shaded brown region and  dot-dashed brown contour.  
	The CMS collaboration has studied a similar process at 13 TeV with an integrated luminosity of $77.3~{\rm fb}^{-1}$ , 
	but required a dilepton resonance from two opposite-sign muons~\cite{Sirunyan:2018nnz}, which leads to the exclusion of the red shaded region.
	}
	\label{fig:independentcoupling}
\end{figure}

We show the parameter space for $\Delta a_{e/\mu}$ in Eq.~(\ref{eq:g-2-e}) and Eq.~(\ref{eq:g-2-mu}) 
in Fig.~\ref{fig:independentcoupling} and the relevant constraints for the couplings are added 
in the plot. 
For the coupling to electrons,  using electron beam, the beam dump experiments E137 \cite{Bjorken:1988as}, 
E141 \cite{Riordan:1987aw}, and Orsay \cite{Davier:1989wz} may produce scalars via Bremsstrahlung-like
process. The scalar would travel macroscopic distances and decay back to electron pairs. The lack of observation
of such events results in the orange shaded exclusion region \cite{Batell:2016ove,Liu:2016qwd} in
Fig.~\ref{fig:independentcoupling} (a). The JLab experiment HPS \cite{Battaglieri:2014hga} projection 
for scalars \cite{Batell:2016ove} is plotted as a  region bounded by the dot-dashed dark cyan line as well. 
 
The BaBar collaboration searches for dark photons through the
process $e^+ e^- \to \gamma A'$ \cite{Lees:2014xha}, where $A' \to \ell^+ \ell^-$ decays democratically.
Ref. \cite{Knapen:2017xzo} recasts the results and give constraints for scalars via $e^+ e^- \to \gamma S$,
which is shown in green shaded region in Fig.~\ref{fig:independentcoupling} (a).  In the BaBar
study, $A'\to \mu^+ \mu^-$ channel is more sensitive than $e^+ e^-$. The constraint for scalar
from \cite{Knapen:2017xzo} applies for ${\rm BR}(S \to \mu^+\mu^-) \gg {\rm BR}(S \to e^+ e^-)$, which
is the case for coupling proportional to lepton mass. If the scalar decays to $e^+e^-$ dominantly,
the limit will be weaker by an order one factor. The process $e^+ e^- \to \gamma S$ at Belle II~\cite{Abe:2010gxa, 
Kou:2018nap} has also been studied to obtain the projected sensitivity \cite{Batell:2016ove}, 
which is plotted as dot-dashed green line in Fig.~\ref{fig:independentcoupling} (a). 
In the lower mass region, the KLOE collaboration provides the constraints for a similar process~\cite{Anastasi:2015qla}, 
and these constraints have been re-interpreted into bounds on the scalar couplings in Ref.~\cite{Alves:2017avw}.

For the coupling to muon, the BaBar collaboration searches the dark photon with muonic coupling via the
$e^+ e^- \to \mu^+ \mu^- A'$ process \cite{TheBABAR:2016rlg}, with $A' \to \mu^+ \mu^-$. It has been
re-casted by the authors of Ref.~\cite{Batell:2016ove, Batell:2017kty} for a scalar with muonic coupling and 
we plotted the  excluded region in Fig.~\ref{fig:independentcoupling} (b) by the shaded green area. 
The future projection for Belle-II \cite{Batell:2017kty, Kou:2018nap} is also shown, 
bounded by the dot-dashed green line. 

At the LHC Run-I, the ATLAS collaboration 
has searched for exotic Z decays, $Z \to 4 \mu$ \cite{Aad:2014wra} with both 7 TeV 
and 8 TeV data. It has been interpreted as a constraint on $Z \to \mu^+ \mu^- S$ by Ref.~\cite{Batell:2017kty}, 
which is shown in Fig.~\ref{fig:independentcoupling} (b) as a  shaded brown region. Ref.~\cite{Batell:2017kty}
has also projected this limit for high luminosity LHC (HL-LHC) and we show it as a region bounded by the dot-dashed 
brown line. Recently, the CMS collaboration has studied the exotic Z decay process $Z \to Z' \mu^+ \mu^-$ at 13 TeV 
with integrated luminosity $77.3~{\rm fb}^{-1}$ \cite{Sirunyan:2018nnz}, which constrained the production
cross-section and exotic Z decay BR($Z \to Z' \mu^+ \mu^-$) as a function of the $Z'$ mass. We recast this
constraint for a scalar which couples to muon and plotted as shaded red region in 
Fig.~\ref{fig:independentcoupling} (b). 
Since the ATLAS search for exotic Z decay $Z \to 4 \mu$ \cite{Aad:2014wra} does not require a dilepton
resonance from the four muon, its HL-LHC projection is weaker than the CMS 13 TeV limit with 
$77.3~{\rm fb}^{-1}$ \cite{Sirunyan:2018nnz}.

For beam dump experiments, whether $\phi_R$ is long-lived is crucial. If $\phi_R$ couples to muons only, 
it can only decay to diphoton when $m_{\phi_R} < 2 m_\mu$ which could be long-lived. The beam dump constraints
could apply in this case due to its small coupling to photons~\cite{Batell:2017kty}. However, in our model,
$\phi_R$ will also couple to electrons with the same coupling strength as $\phi_I$. Therefore, 
the beam dump constraints do not apply for $\phi_R$ under the assumption that it is heavier than $\phi_I$.

We only plotted the relevant limits for the EFT model in Fig.~\ref{fig:independentcoupling}. 
For readers who are interested in more detailed future sensitivity projections and new proposals 
from beam dump, collider searches and cosmology constraints for light scalar coupled to 
leptons, they can be found in Refs.~\cite{Batell:2016ove, Knapen:2017xzo, Batell:2017kty} and references therein.

\section{ EFT model with a light complex scalar}
\label{sec:EFTModel}

In this section, we demonstrate at the effective field theory (EFT)  level that a complex scalar $\phi$, accompanied 
with some symmetry assumption can simultaneously solve the $\Delta a_{e}$ and $\Delta a_\mu$ anomalies. The 
gauge charge of $\phi$ and the global $U(1)_{\rm PQ}^{e}$ charges are presented  in Table.~\ref{tab:gaugecharge}. 

\begin{table}[h]
\begin{tabular}{|c||c|c||c||}
\hline 
filed  &$SU(2)_L$  &$U(1)_Y$ & $U(1)_{\rm PQ}^{e}$ \\  \hline
\hline
 $H$    & 2  & $\frac{1}{2}$  & 0   \\
 \hline
 $\phi$     & 1  &  0                  &  -2    \\
\hline
$L_e$     & 2   & $\frac{1}{2}$    & 1   \\
\hline
$e_R$     & 1   & -1    & -1   \\
\hline
\end{tabular}
\caption{All particles with $ SU(2)_L \times U(1)_Y \times U(1)_{\rm PQ}^{e}$ charge
specified, where $U(1)_{\rm PQ}^{e}$ is a global Peccei-Quinn-like symmetry.
$H$ and $L_e$ ($e_R$) are SM Higgs and left-handed (right-handed) electron, 
while $\phi$ is the new light singlet complex scalar.
}
 \label{tab:gaugecharge}
\end{table}

Given the particle content and charge in Table.~\ref{tab:gaugecharge}, we can write down
the effective theory Lagrangian as
\begin{align}
\mathcal{L}_{\rm EFT} =  \frac{\phi^*}{\Lambda_e} \bar{L}_e H e_R + y_\mu \bar{L}_\mu H \mu_R
+   \frac{\phi^* \phi}{\Lambda_\mu^2}  \bar{L}_\mu H \mu_R + H.c.  ,
\label{eq:EFT-Lag}
\end{align}
where $\Lambda_{e,\mu}$ are interaction scales, $H$ is the SM Higgs, $L_{e,\mu}$ are 
SM left-handed doublets for leptons and $e_R, ~ \mu_R$ are the right-handed SM leptons.
In principle, the tau leptons could also appear in the last two terms in Eq.~(\ref{eq:EFT-Lag}),
thus flavor violation coupling can be generated. We postpone the discussion of this issue to 
section~\ref{sec:discussionElecMuon}. Both the SM Higgs and the new scalar $\phi $ can get 
vacuum expectation values (vevs),
\begin{align}
H = \frac{1}{\sqrt{2}} \left(v + h + i G^0 \right), \quad 
\phi = \frac{1}{\sqrt{2}} \left(v_\phi + \phi_R + i \phi_I \right) .
\end{align}

For the electron, its mass can only come from the first term which is a dimension 5 operator, 
while the muon mass can come from the second and third term. It is straight forward to obtain
the following relations
\begin{align}
& m_e = \frac{v v_\phi}{2 \Lambda_e}, \quad m_{\mu} = \frac{y_\mu v}{\sqrt{2}} 
+ \frac{v v_\phi^2}{2\sqrt{2} \Lambda_{ \mu}^2} ,
\label{eq:masses} \\
&g_{\phi_{R}}^{e, {\rm EFT}} = - g_{\phi_{I}}^{e, {\rm EFT}} = \frac{v}{2 \Lambda_e}=\frac{m_e}{v_\phi} , 
\quad g_{\phi_{R}}^{\mu, {\rm EFT}} = \frac{v_\phi v}{\sqrt{2} \Lambda_\mu^2} .
\label{eq:gIEFT}
\end{align}
We find that the CP-odd $\phi_I$ and CP-even scalars $\phi_R$ couples to electron with the same strength.
For the electron anomalous magnetic dipole, the  contributions  from the two scalars have opposite signs. To obtain
negative $\Delta a_{e}$, the $\phi_I$ contribution has to be larger than the $\phi_R$ one, which can be satisfied
by requiring $m_{\phi_I} \ll m_{\phi_R}$. We emphasize that such requirement is natural to achieve,
because if $U(1)_{\rm PQ}^{e}$ is spontaneously broken, the Goldstone $\phi_I$ is massless.
However, we have to downgrade the continuous global symmetry to a discrete one, for example,
adding a soft breaking term, e.g. $\mu_4^2 \phi_I^2$ term to give mass to $\phi_I$. It can also get
mass from hidden confinement scale \cite{Marques-Tavares:2018cwm}. The mass
of $\phi_R$ is not dictated by symmetry breaking, thus can be larger.

In the EFT model, we have 6 free parameters, $\Lambda_e$, $\Lambda_\mu$, $y_\mu$, 
$v_\phi$, $m_{\phi_I}$ and $m_{\phi_R}$.  With the electron and muon masses, we can eliminate
$\Lambda_e$ and $y_\mu$. To fit the anomalous magnetic moment $\Delta a_{e}$, we further
eliminate $v_\phi$. From the electron sector, only $m_{\phi_I}$ is a free parameter, though is
limited to a small range $10-100$ MeV from the constraints in Fig.~\ref{fig:independentcoupling} (a).
We choose $m_{\phi_I} \sim 15$ MeV as our benchmark, which also implies $g_{\phi_I}^e \sim 10^{-4}$. 
Let us stress that for $\Delta a_{e}$, the 1-loop \cite{PhysRevLett.65.21} correction is suppressed 
by the electron mass, and hence 
the 2-loop Barr-Zee diagram could be dominant if $\phi_{I}$ couples to other heavy charged fermions 
\cite{Giudice:2012ms, Davoudiasl:2018fbb}.
In our case, however, the $\phi_I$ only couples to the electron due to the PQ charge assignment and thus
the 2-loop contribution is much smaller than the 1-loop one~\cite{TuckerSmith:2010ra}.

\begin{figure}[htb]
	\includegraphics[width=0.325 \columnwidth]{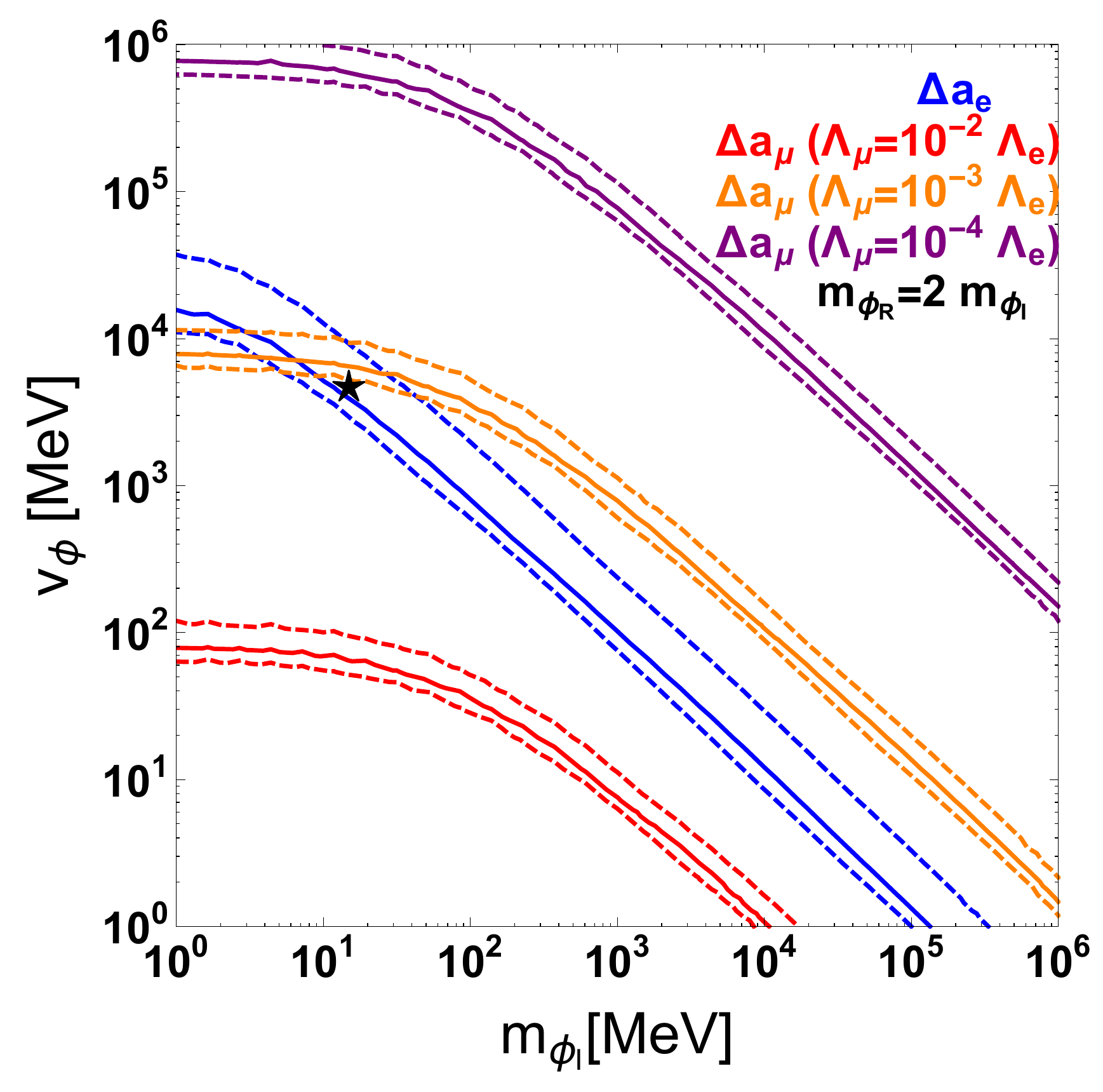} 
	\includegraphics[width=0.325 \columnwidth]{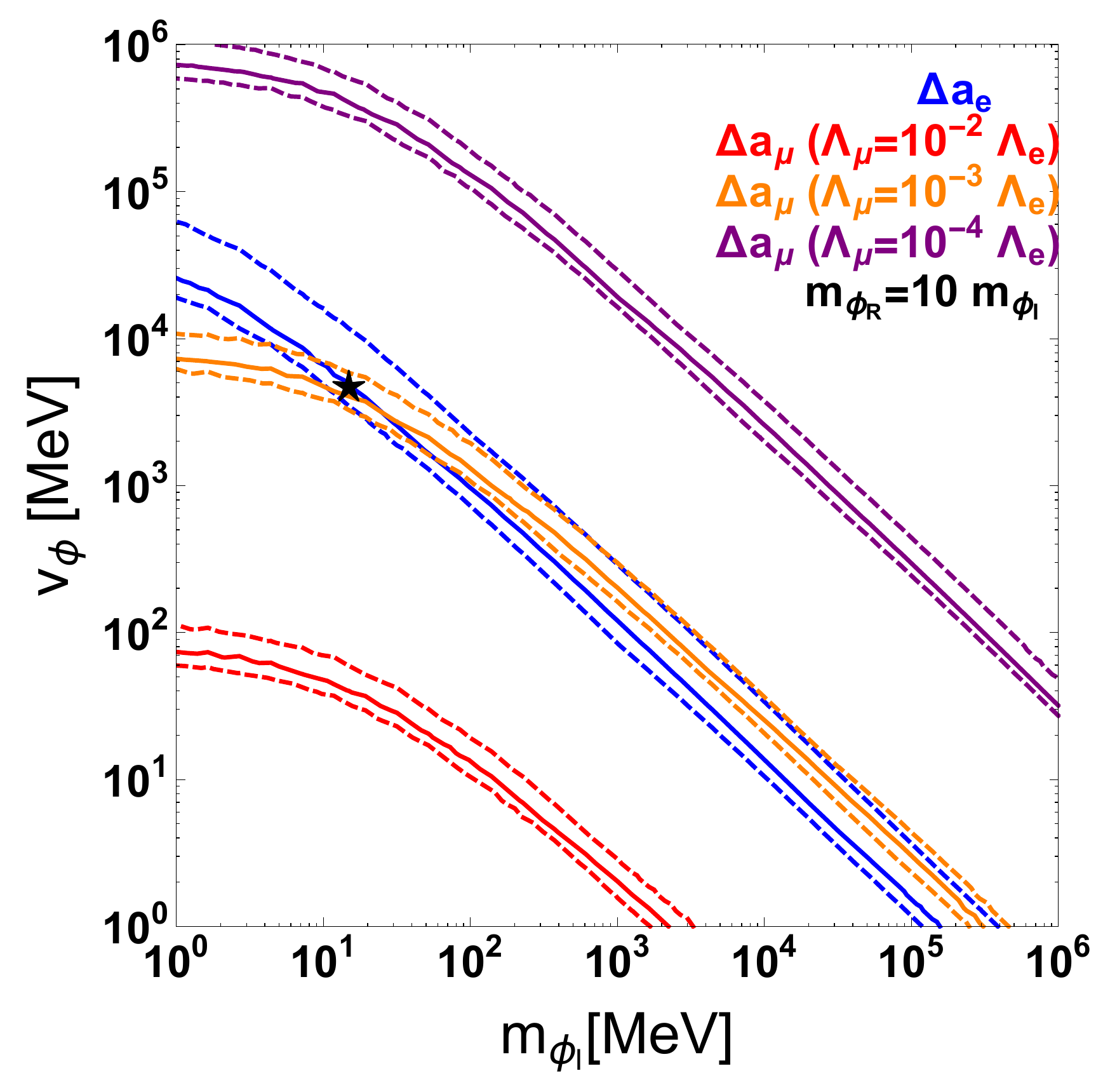} 
	\includegraphics[width=0.325 \columnwidth]{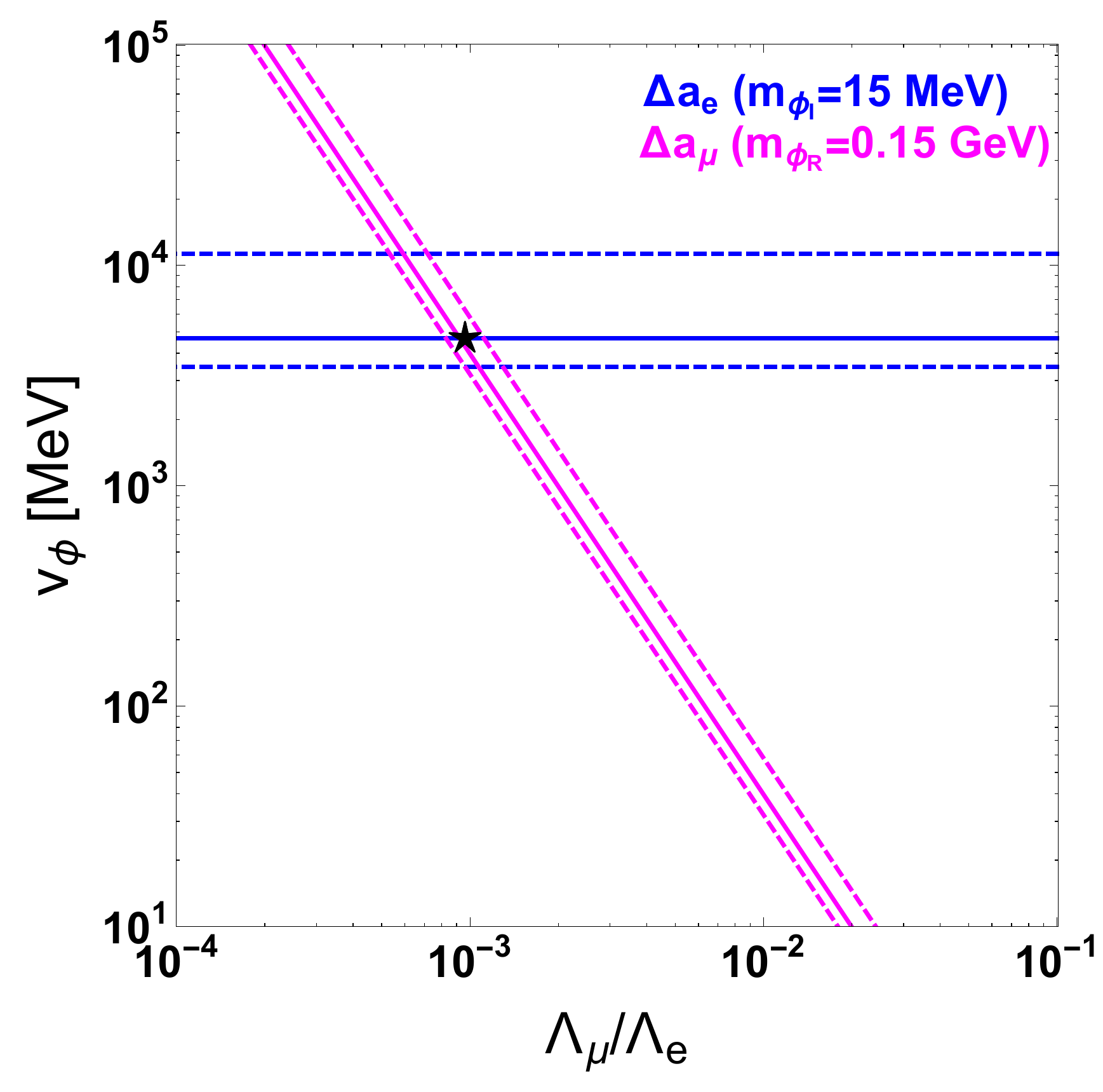} 
	\caption{
		The EFT parameter space with parameters $v_{\phi}$, $m_{\phi_I}$, $m_{\phi_R}$, 
		and $\Lambda_{ \mu}$ for $\Delta a_{e/\mu}$ anomalies.
	}
	\label{fig:EFT-fit}
\end{figure}

The $\Delta a_{\mu}$ defines a band in $\Lambda_\mu$ and $m_{\phi_R}$ region as well. 
As a result, after applying two lepton mass and $\Delta a_{e / \mu}$ requirements, we are left 
with 2 degree of freedom (d.o.f.) as $m_{\phi_{I}}$ and $m_{\phi_{R}}$. We list a benchmark point with 
$m_{\phi_R} \sim 15$ MeV and $m_{\phi_R} \sim 0.15$ GeV as an example in Table.~\ref{tab:benchmark1}. 
In Fig.~\ref{fig:EFT-fit}, we show the fits for $\Delta a_{e/\mu}$ anomalies with the parameters $v_{\phi}$, 
$m_{\phi_I}$, $m_{\phi_R}$, and $\Lambda_{ \mu}$.

\begin{table}[htb]
	\begin{tabular}{|c|c|c|c|c|}
		\hline 
		$v_{\phi}$ (GeV) & $m_{\phi_I}$ (MeV) & $\Lambda_e$ (GeV) &  $\Lambda_{ \mu}$ (GeV) &
		$m_{\phi_R}$ (GeV )  \\  
		\hline
		$4.7 $      & 15   & $1.12 \times 10^6$   & $1080$  & 0.15 \\
		\hline
	\end{tabular}
	\caption{The benchmark for EFT model. The parameter $y_\mu$ is determined by muon
	mass which is not listed here. The EFT model has 2 d.o.f., $m_{\phi_{I}}$ and $m_{\phi_{R}}$, after applying
	all the constraints and signal requirements. The change of $m_{\phi_R}$ only affects $\Lambda_{ \mu}$,
	while $v_\phi$ and $\Lambda_e$ are already fixed by the electron mass and $\Delta a_e$. $m_{\phi_I}$ is
	limited to a small range $10-100$ MeV by relevant constraints. This benchmark is labeled as a black star in Fig.~\ref{fig:independentcoupling}.
	}
	\label{tab:benchmark1}
\end{table}

In the EFT model, we further consider the possibility that the muon mass comes  from the
dimension 6 operator, e.g. when $y_\mu =0$. In this case, $\Lambda_{ \mu} =  135$ GeV is 
enforced by the muon mass. It implies that $g_{\phi_{R}}^{\mu, {\rm EFT}} \approx 0.045$ and 
the $\phi_R$ mass is around $26 - 50$ GeV. In this case, there is no free parameter left in the 
EFT model.  This possibility is constrained by the recent analysis of the CMS 13 TeV data with 
$77.3~{\rm fb}^{-1}$ \cite{Sirunyan:2018nnz}  shown in Fig.~\ref{fig:independentcoupling} (b),
that restrict $\phi_R$ masses smaller than 38.5 GeV is excluded.  Although masses of the
order of 40~GeV would be allowed, leading to values of $a_\mu$ which deviate by less
than 1~$\sigma$ from the experimental value, 
one more issue with this region of parameters is that $\Lambda_{ \mu}$ is around 135 GeV, which implies
new physics should be much lighter than in the original benchmark. We leave the exploration of this 
parameter space for future work.

\section{UV complete model with a light complex scalar}
\label{sec:UVModel}

In this section, we show the UV completion of the EFT Lagrangian in Eq.~(\ref{eq:EFT-Lag}). 
The particle content of the UV model is listed in Table~\ref{tab:Higgsdoublet}.
It contains three Higgs doublet $\Phi_{1,2,3}$, where $\Phi_2$ will become the SM-like Higgs.
A SM singlet complex scalar $\phi$ transforms under an approximate $U(1)$ PQ-like symmetry, 
while $\Phi_1$, $L_e$ and $e_R$ also transform under it. The symmetry has to be softly broken 
to allow a massive $\phi_I$. $\Phi_{2,3}$ have no global charge assigned.

\begin{table}[htb]
	\begin{tabular}{|c||c|c||c||}
		\hline 
		filed  &$SU(2)_L$  &$U(1)_Y$ & $U(1)_{\rm PQ}^{e}$ \\  \hline
		\hline
		 $\Phi_1$  &  2   &  $\frac{1}{2}$ & 2  \\
		 \hline
		$\Phi_2$  & 2  & $\frac{1}{2}$  & 0    \\
		\hline
		$\Phi_3$  & 2  & $\frac{1}{2}$  & 0    \\
		\hline
		$\phi$     & 1  &  0                  &  -2    \\
		\hline
		$L_e$  & 2  & $\frac{1}{2}$  & 1   \\
		\hline
		$e_R$  & 1  & $-1$  & -1   \\
		\hline
	\end{tabular}
	\caption{The particles under $ SU(2)_L \times U(1)_Y \times U(1)_{\rm PQ}^{e}$
		, where $U(1)_{\rm PQ}^{e}$ is a global Peccei-Quinn-like symmetry. 
		The Higgs doublet $\Phi_1$ and $\Phi_3$ are supposed to be
		heavy degrees of freedom, which are integrated out in the effective theory. 
		The mixing between the scalars are assumed to be small and 
		$\Phi_2$ will be the SM-like Higgs.
	}
	\label{tab:Higgsdoublet}
\end{table}

\begin{figure}[htb]
	\begin{tabular}{ccc}
		\includegraphics[width=0.25 \columnwidth]{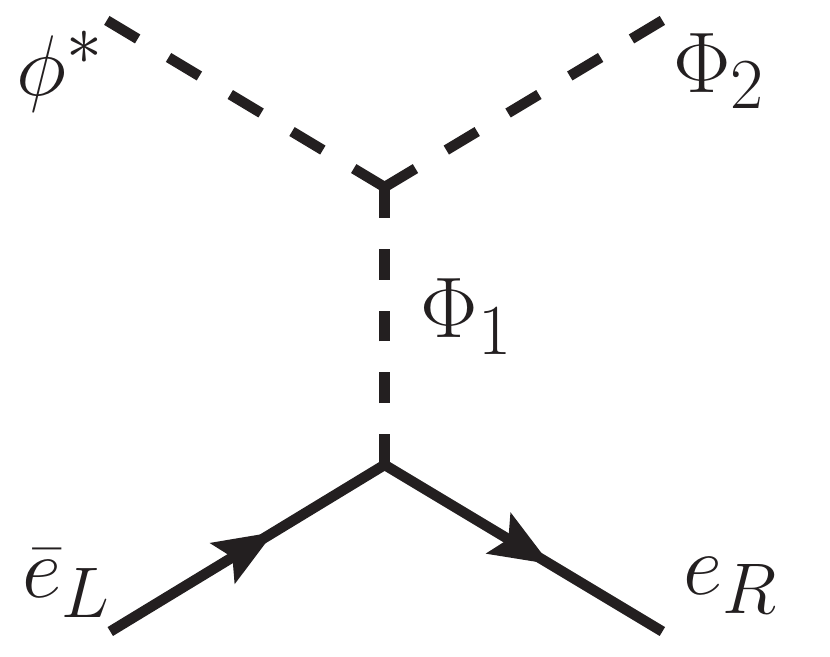} &
		\includegraphics[width=0.23 \columnwidth]{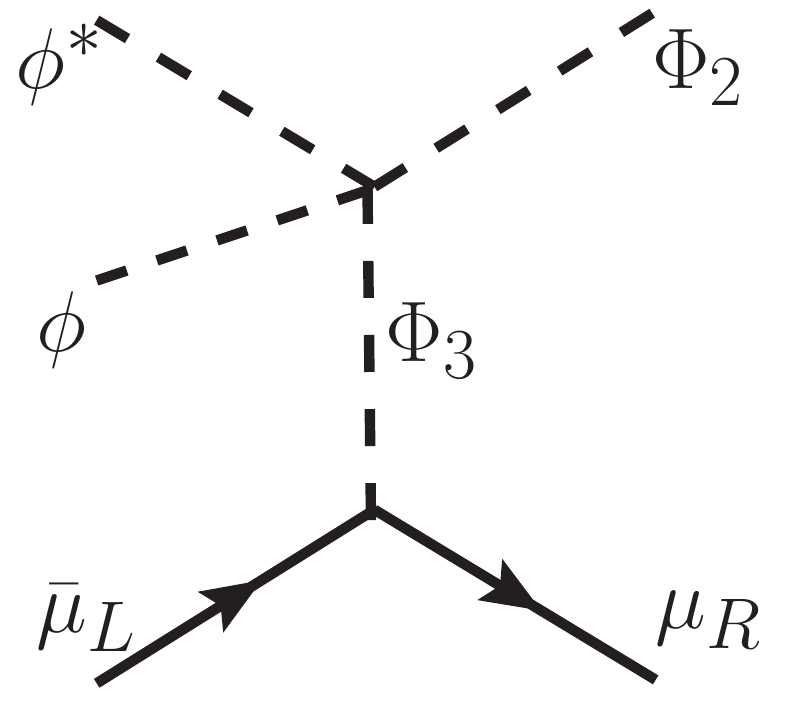} &
		\includegraphics[width=0.38 \columnwidth]{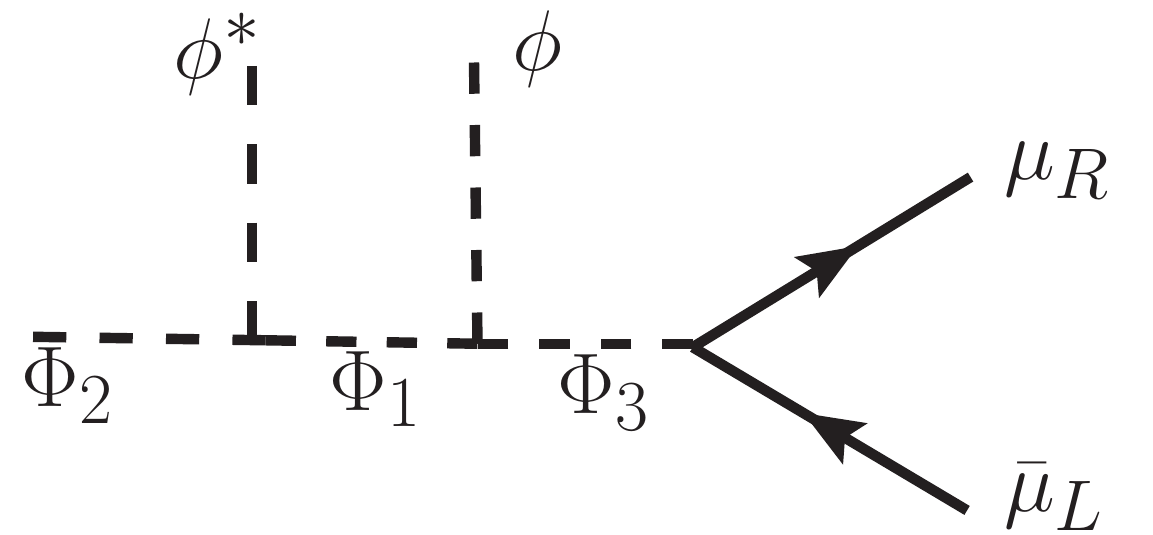} \\
		(a) & (b) & (c)
	\end{tabular}
	\caption{
		The relevant Feynman diagrams for generating EFT Lagrangian. The figure (a) is responsible
		for $ \phi^*\bar{L}_e H e_R + H.c.$, while (b) and (c) are responsible for 
		$ \phi^*\phi (\bar{L}_\mu H \mu_R + H.c. )$.
	}
	\label{fig:EFTgeneration}
\end{figure}

\subsection{The electron sector}
We need the Higgs doublet $\Phi_1$ charged under $U(1)_{\rm PQ}^{e}$ to generate the
dimension 5 operator in the EFT Lagrangian which is responsible for the electron mass. The
relevant Feynman diagram is shown in Fig.~\ref{fig:EFTgeneration} (a), where the
heavy $\Phi_1$ is integrated out. The relevant UV Lagrangian for the electron sector is given by,
\begin{align}
\mathcal{L}_{\rm UV}^e & = V(\Phi_1, \Phi_2)_{\rm 2HDM}^{U(1)} +\left( y_e \bar{L}_e \Phi_1 e_R + H.c.\right) 
\nonumber \\
&+ V(\phi) + \frac{1}{2} \mu_4^2 \phi_I^2 \nonumber \\
& + (\phi^* \phi) \left( \lambda_5 \Phi^\dagger_1 \Phi_1
+  \lambda_6 \Phi^\dagger_2 \Phi_2 \right) 
+ \mu_8 \left(\Phi^\dagger_1 \Phi_2 \phi^* + H.c. \right) .
\label{eq:LagPhi1}
\end{align}
After getting a vev, the neutral component in each of the Higgs doublets is 
\begin{align}
 \Phi_j^0 =  \frac{1}{\sqrt{2}} \left(v_j + h_j + i a_j \right),
\end{align}
where we assume $v_3 \ll v_1 \ll v_2$. For further simplicity, we assume the alignment limit
that $\Phi_2 \approx H$ and the mixing angles between $\Phi_i$ and $\phi$ are small.
We neglect $\Phi_3$ at this moment, since it is not necessary for generating the EFT operators 
in the electron sector.

In Eq.~(\ref{eq:LagPhi1}), the coefficients are all real, as required by CP conservation.
In the first line, the scalar potential $V(\Phi_1, \Phi_2)_{\rm 2HDM}^{U(1)}$ \cite{Branco:2011iw} 
is the usual two Higgs doublet model (2HDM) 
potential subject to the global $U(1)$ charge.  The Yukawa coupling for the electron is mediated
by $\Phi_1$ only. In the second line, the singlet scalar potential $ V(\phi)$ contains the quadratic 
$\phi^* \phi$ and quartic $(\phi^* \phi)^2$ terms satisfying the global $U(1)_{\rm PQ}^{e}$ symmetry.
However, we explicitly add the $\mu_4^2 \phi_I^2$ term to break $U(1)_{\rm PQ}^{e}$ softly, since
otherwise $\phi_I$  will be a massless pseudo-Goldstone boson.
In the third line \footnote{It is termed as leptonic Higgs portal in \cite{Batell:2016ove}, 
where a real singlet scalar example is demonstrated.}, 
the $\mu_8$ term is special because it contributes to the splitting of the mass for CP-odd 
scalars with respect to the CP-even ones.  Regarding the CP-odd sector, the mass
eigenstates are a heavy massive $A^0$, a Goldstone boson $G^0$  eaten by Z gauge boson
and a remaining pseudo-Goldstone $\phi_I^{'}$ for the global $U(1)_{\rm PQ}^{e}$. 
In the small mixing setup, the mass eigenstates $A^0$, $G^0$ and $\phi_I^{'}$ are mostly 
$a_1$, $a_2$ and $\phi_I$, respectively.

Following \cite{Liu:2017xmc}, the mass for $A^0$ and $\phi_I^{'}$ and their mixing between
different states are given by
\begin{align}
& m_{A^0}^2 =  - \mu_8 v_2 \frac{v_1^2 + v_\phi^2}{\sqrt{2} v_1 v_\phi}  ,  
\quad m_{\phi_I^{'}}^2 =  \mu_4^2 \frac{v_\phi^2}{v_1^2 + v_\phi^2} , 
\label{eq:CPoddmass}\\
& a_1 = \frac{v_\phi}{\sqrt{v_1^2 + v_\phi^2}} A^0 + \frac{v_1}{v} G^0 
- \frac{v_1}{\sqrt{v_1^2 + v_\phi^2}} \phi_I^{'} +\mathcal{O}\left( \frac{\mu_4^2}{\mu_8 v_2}\right) 
\label{eq:CPoddmixing1},\\
& \phi_I = \frac{v_1}{\sqrt{v_1^2 + v_\phi^2}} A^0 + 0 \times G^0 
+ \frac{v_\phi}{\sqrt{v_1^2 + v_\phi^2}} \phi_I^{'} +\mathcal{O}\left( \frac{\mu_4^2}{\mu_8 v_2}\right),
\label{eq:massmixing}
\end{align}
where $v \equiv \sqrt{v_1^2 + v_2^2}$ and we have taken only the leading term 
under assumption $v_2 \gg  v_{\phi}, v_1$. If we further impose
$v_\phi \gg v_1$, then our assumption that scalar mixing is small can be satisfied.
From the mixing in the UV model, we can calculate the coupling $g_{\phi_I}^e$ that
\begin{align}
g_{\phi_I}^{e, {\rm UV}} = - \frac{y_e}{\sqrt{2}} \frac{v_1}{\sqrt{v_1^2 + v_\phi^2}} =
- \frac{m_e}{\sqrt{v_1^2 + v_\phi^2}} .
\label{eq:gphiIeUV}
\end{align}
After integrating out $\Phi_1$, one can also obtain the interaction scale $\Lambda_e$ that
\begin{align}
\frac{1}{\Lambda_e} =  y_e \frac{\mu_8}{ m_{A_0}^2} . \label{eq:Lambdae}
\end{align}
In Eq.~(\ref{eq:Lambdae}), due to CP conservation, the integrated particle should be 
the CP-odd component in $\Phi_1$, thus the denominator is the mass of $A_0$ squared. 
Applying Eq.~(\ref{eq:CPoddmass}) and Eq. (\ref{eq:Lambdae}), one can check that 
$g_{\phi_I}^{e, {\rm EFT}} $ in Eq.~(\ref{eq:gIEFT}) agrees with $g_{\phi_I}^{e, {\rm UV}} $.
One can also see that the mass of $A^0$ can be easily as large as 1~TeV 
if $\mu_8$ is electroweak scale and $v_\phi/v_1$ is large.

\subsection{The muon sector} 

In this section, we describe the UV model which can generate the dimension 6 operator in
$\mathcal{L}_{\rm EFT}$, which is responsible for the $\phi_R$ coupling to muons.
A third Higgs doublet $\Phi_3$ is essential and it has to carry the same quantum charge 
as SM-like Higgs $\Phi_2$. The relevant Lagrangian is 
\begin{align}
\mathcal{L}_{\rm UV}^{\mu} & = V(\Phi_2, \Phi_3)_{\rm 2HDM} 
+  (y_{\mu} \bar{L}_\mu \Phi_2 \mu_R + y_{\mu 3} \bar{L}_\mu \Phi_3 \mu_R + H.c.)  \nonumber \\
& + V(\phi) + (\phi^* \phi) \left( \lambda_6 \Phi^\dagger_2 \Phi_2+  \lambda_8 \Phi^\dagger_3 \Phi_3 \right) \nonumber \\
& + \lambda_9 (\phi^* \phi) \left(  \Phi^\dagger_2 \Phi_3 +   H.c. \right)
+ \mu_9 \left(\Phi^\dagger_1 \Phi_3 \phi^* + H.c. \right)  ,
\label{eq:LagPhi3}
\end{align}
where the coefficients are real. 

The first line in Eq.~(\ref{eq:LagPhi3}) contains a general  2HDM  scalar potential $V(\Phi_2, \Phi_3)_{\rm 2HDM}$.
The last two terms in that line are the Yukawa couplings for the muon. We will again assume hierarchical vevs, 
$v_3 \ll v_1 \ll v_\phi \ll v_2$, so that the muon mass predominantly comes from $\Phi_2$ and $y_{\mu 3}$ is free
from the muon mass constraint.
The second line contains the scalar potential for $\phi$ and the quartic coupling between $\phi$ and $\Phi_{2,3}$.
Since $v_3\sim 0$, if  we require $\lambda_6 \ll 1$, the quartic term in the second line does not
induce a  large mixing between the different scalars\footnote{As it is discussed in Appendix \ref{sec:CPoddUV}, 
the presence of large $\lambda_6$ or $\mu_8$ term combined with large $h_2$-$h_3$ mixing from $V(\Phi_2, \Phi_3)_{\rm 2HDM} $
can lead to relevant contributions to the dimension 6 operator at low energy. }.
Since $\Phi_2$ and $\Phi_3$ have the same quantum numbers, the potential $V(\Phi_2,\Phi_3)_{\rm 2HDM}$ 
may include  a quadratic term $m_{23}^2 \Phi_3^\dagger \Phi_2 + H.c.$, 
while the third line contains the term proportional to $\lambda_9$ which may also lead to a similar term when $\phi$ 
acquires a vev. These two terms contribute to the dimension 4 and 6 operators responsible for the muon mass 
and the coupling of $\phi_R$ to the muons in the effective field theory described by
$\mathcal{L}_{\rm EFT}$, Eq.~(\ref{eq:EFT-Lag}).   Finally, the term proportional to the trilinear mass parameter $\mu_9$, 
in combination with the $\mu_8$-induced interactions, can also contribute to the $\phi_R$ coupling to muons. Although 
all these contributions may coexist, we shall treat them in a separate way for simplicity of presentation. 

\subsubsection{Generating the operators from quartic scalar  interactions}

The term proportional to the $\lambda_9$ coupling in Eq.~(\ref{eq:LagPhi3}) can generate the Feynman diagram depicted in 
Fig.~\ref{fig:EFTgeneration} (b), which  can lead, after integrating out $\Phi_3$, to the coupling of $\phi_R$ to muons in
the EFT, Eq.~(\ref{eq:EFT-Lag}). This coupling is given by
\begin{align}
g_{\phi_R}^{\mu, {\rm EFT}} = \frac{v_\phi v}{\sqrt{2} \Lambda_{ \mu}^2} = 
y_{\mu 3} \lambda_9 \frac{v_\phi v_2}{\sqrt{2} m_{h_3}^2} ,  \label{eq:gEFTmuphiR4point}
\end{align}
where $m_{h_3}^2$ is the CP-even scalar mass from $\Phi_3$. The interaction scale $\Lambda_{ \mu}$
is related to the heavy Higgs parameters by the relation
\begin{align}
\frac{1}{\Lambda_{ \mu}^2} = \frac{y_{\mu 3} \lambda_9}{m_{h_3}^2}
\label{eq:mh3case1} .
\end{align}

Given the fact that the $\lambda_{9}$ term gives the off-diagonal mass terms between $\phi_R$ 
and $h_3$, we can calculate the mass matrix and obtain the mixing angle,
\begin{align}
& M^2_{\phi_R h_3}  = 
\left(\begin{array}{cc}
m_{\phi_R}^2 & \lambda_9 v_\phi v_2 
\\
\lambda_9 v_\phi v_2  & m_{h_3}^2
\end{array}\right) , \\
& \sin \theta_{\phi_R h_3}  \approx  \lambda_9 \frac{v_\phi v_2 }{m_{h_3}^2} . \label{eq:quarticMIX}
\end{align}
Assuming that  $h_3$ and $\phi_R$ only have a small mixing between themselves ($\sin \theta_{\phi_R h_3} \ll 1$) 
and negligible mixing with other fields, the coupling between $\phi_R$ and the muon from the UV model is 
\begin{align}
g_{\phi_R}^{\mu, {\rm UV}} = \frac{y_{\mu 3}}{\sqrt{2}} \sin \theta_{\phi_R h_3} .
\end{align}
One can easily check that it agrees with $g_{\phi_R}^{\mu, {\rm EFT}}$ in Eq.~(\ref{eq:gIEFT})
and Eq.~(\ref{eq:gEFTmuphiR4point}).

In the UV model, the $\lambda_9$ and $\lambda_8$ terms in $\mathcal{L}_{\rm UV}^{\mu}$ 
contain only $\phi^* \phi$, thus $\phi_I$ couples to
muon only in quadrature and can not contribute to $\Delta a_{\mu}$. 
Given that $\Lambda_{ \mu}$ needs to be about $1080$ GeV (see Table~\ref{tab:benchmark1}), the scalar boson 
$h_3$ can be easily  
heavier than $\mathcal{O}(1)$ TeV, as can be seen from Eq.~(\ref{eq:mh3case1}).

\subsubsection{Generating the operators from triplet scalar interaction}
\label{sec:MuCouplingfromTriplet}
We can generate the CP-even scalar $\phi_R$ coupling to muon via Fig.\ref{fig:EFTgeneration} (c),
after integrating out the heavy $h_1$ and $h_3$ scalar bosons. As emphasized above, it requires  the simultaneous action of the two triple scalar couplings $\mu_9 \Phi^\dagger_1
\Phi_3 \phi^*$ and $\mu_8 \Phi^\dagger_1 \Phi_2 \phi^*$. According to \cite{Liu:2017xmc}, under the assumption $v_1 \ll v_\phi \ll v_2$, $m_{h_1}^2$ is the same order as $m_{A_0}^2$ in Eq.~(\ref{eq:CPoddmass}), what is also confirmed in the
full UV model calculation presented in Appendix \ref{sec:CPoddUV}.
The EFT coupling between $\phi_R$ and muon can be 
computed as 
\begin{align}
g_{\phi_R}^{\mu, {\rm EFT}} =
\frac{y_{\mu 3} v_\phi v_2 \mu_8 \mu_9}{\sqrt{2} m_{h_1}^2 m_{h_3}^2} \approx 
y_{\mu 3} \frac{v_1 \mu_9}{m_{h_3}^2}, \label{eq:gEFTmuphiR}
\end{align}
where $m_{h_{1, 3}}^2$ are the CP-even scalar mass from $\Phi_{1,3}$. The interaction scale $\Lambda_{ \mu}$ in this case is 
\begin{align}
\frac{1}{\Lambda_{ \mu}^2} = \frac{y_{\mu 3} \mu_8 \mu_9}{m_{h_1}^2 m_{h_3}^2} .
\end{align}

In the UV model, the $\phi_R$ coupling to muon again comes from mixing with $h_3$. 
We calculate the mass matrix and obtain the mixing angle via $\mu_9 \Phi^\dagger_1
\Phi_3 \phi^*$ term,
\begin{align}
& M^2_{\phi_R h_3}  = 
\left(\begin{array}{cc}
m_{\phi_R}^2 & \mu_9 v_1
\\
\mu_9 v_1  & m_{h_3}^2
\end{array}\right) , \\
& \sin \theta_{\phi_R h_3}  \approx   \frac{\mu_9 v_1 }{m_{h_3}^2}, \label{eq:MixphiRh3}
\end{align}
where again we  find that $g_{\phi_R}^{\mu, {\rm UV}} \equiv y_{\mu 3}  \sin \theta_{\phi_R h_3} $ 
agrees with $g_{\phi_R}^{\mu, {\rm EFT}}$ again.
In the above discussion, we did not include off-diagonal terms with $h_{1,2}$.  The 
$\phi_R$-$h_3$ mixing, may be modified through the mixing with them.
We did the full calculation in the 3HDM plus a singlet complex scalar in Appendix~\ref{sec:CPoddUV}.
The result contains more terms than Eq.~(\ref{eq:MixphiRh3}), but one can tune down some parameters 
to converge to this result, while keeping the $\Phi_{1,3}$ scalars heavy. Such tuning is also in agreement of 
the initial assumption that the mixing between different scalars is small, see Appendix~\ref{sec:CPoddUV}.

From the benchmark point,  we can see that a coupling $g_{\phi_R}^{\mu, {\rm EFT}} \sim 0.7 \times 10^{-3}$
can fit the $\Delta a_\mu$ anomaly. One can infer the mass square $m_{h_3}^2 \simeq 10^3 y_{\mu 3} v_1
\mu_9$ from Eq.~(\ref{eq:gEFTmuphiR}). 
With a large $\mu_9 \simeq \mathcal{O}({\rm few})$~TeV and $v_1 \sim 1$ GeV, $h_3$ mass can be  larger than 
$\mathcal{O}(1)$ TeV.
 
Given the fact that the $\phi_I^{'}$ mass is much smaller than $\phi_R$, any small mixing between $\phi_I^{'}$ and the
CP-odd components of $\Phi_2$ and $\Phi_3$, would induce a coupling of $\phi_I^{'}$ to muons, that could make
the contribution from $\phi_I^{'}$ to $\Delta a_\mu$  larger than the one of  $\phi_R$.  However, 
 in the full calculation within the 3HDM plus singlet
scalar potential, presented in Appendix~\ref{sec:CPoddUV}, the mass eigenstate $\phi_I^{'}$ 
only mixes with $a_1$ in the leading order of $v_1/v_\phi$.  In fact, the absence of mixing with $a_{2,3}$ can
be simply understood from the pseudo-Goldstone nature of this particle.  Thus,  the components of $\phi_I^{'}$ 
are approximately described by Eq.~(\ref{eq:CPoddmixing1}), and $\phi_I^{'}$ only couples to electrons, as occurs in the EFT model.

\section{Phenomenology constraints}
\label{sec:discussionElecMuon}

There are several important phenomenological constraints to address, once moving from EFT model to the UV model.

\subsection{Heavy scalars and anomalous magnetic moments}

One relevant constraint is the contribution of the heavy scalars to the anomalous magnetic moments.
Although these scalars are integrated out in the EFT, they may contribute in a relevant way.
At large mass, the CP-even scalar contribution to the lepton g-2 is approximately given by
$\Delta a_{\ell}^{\rm even} \approx g_{S \ell}^2/(8 \pi^2) m_{\ell}^2/m_{S}^2 ( \log(m_{S}^2/m_{\ell}^2) -7/6)$, 
while CP-odd scalar contributes as $\Delta a_{\ell}^{\rm odd} \approx g_{S \ell}^2/(8 \pi^2) 
m_{\ell}^2/m_{S}^2 (-  \log(m_{S}^2/m_{\ell}^2) +11/6 )$~\cite{Giudice:2012ms}. 
Neglecting the mild dependence on  log terms,
the anomalous magnetic moments are hence proportional to $g_{\ell}^2 / m_S^2$. 

In the UV model, the light
scalar couples to leptons via mixing with the heavy ones for the pseudo-scalar case, where the mixing angles are related
to vevs due to pseudo-Goldstone nature. Therefore, for light scalar contribution dominating over
the heavy scalar one, the relation  $\sin^2\theta > m_{\rm light}^2 / m_{\rm heavy}^2$ must be satisfied, where
$\sin\theta$ is the mixing angle, while $m_{\rm light}$ and $m_{\rm heavy}$ are the light and heavy scalar masses. 
The mixing angles do not significantly depend on  the mass of the light scalars, $m_{\rm light}$, thus one can always tune down
light scalar mass to meet the requirement. It is easy to find that the $a_1$ contributions to the 
$e$ anomalous magnetic moments is sub-dominant than the $\phi_I$ ones due to the small values of the 
lightest pseudo-scalar mass, while  satisfying the benchmark requirements. 

However,
for CP-even scalar $h_3$ mixing, the mixing angle $\sin\theta$ is proportional to $m_{h_3}^{-2}$.
Therefore, $\sin^2\theta \propto m_{h_3}^{-4}$ and the $\sin^2\theta > m_{\rm light}^2 / m_{\rm heavy}^2$ condition actually
provides an upper bound on the $h_3$ mass. If we choose the benchmark presented in Table~\ref{tab:benchmark1}, 
with $v_\phi = 4.7$ GeV, $m_{\phi_R} =0.15$~GeV, and $g_{\phi_R}^{\mu, \text{EFT}} = 0.7 \times 10^{-3}$, 
for the cases in which the effective low energy couplings are induced by quartic (triplet) scalar interactions, the $h_3$ contribution 
would be smaller than $\phi_R$ provided that 
\begin{equation}
m_{h_3} < \lambda_9 \times 7.7~{\rm TeV}~ \; (m_{h_3} < 6.7 \times \mu_9 v_1 \text{GeV}^{-1}).
\label{eq:upperbmh3}
\end{equation}
To satisfy $g_{\phi_R}^{\mu, \text{EFT}} = 0.7 \times 10^{-3}$, for quartic (triplet) scalar interactions,
one should further demand that $m_{h_3} = 1~ \text{TeV} \times \sqrt{\lambda_9 y_{\mu3}}$ 
($m_{h_3} = 37.8 \sqrt{y_{\mu3} \mu_9 v_1}$), as can be seen from Eqs. (\ref{eq:gEFTmuphiR4point}) and
(\ref{eq:gEFTmuphiR}). 
These requirements can be achieved easily, with $\lambda_9 \sim 1$
and $y_{\mu 3} \sim 1$ for quartic case, while $\mu_9 \sim 1$ TeV, $v_1 \sim 1$ GeV and $y_{\mu 3} \sim 1$
for triplet case. It is worth mentioning that $m_{h_3}$ is about 1~TeV in both cases. Therefore, 
we conclude that in the UV model, under the hierarchical vevs and heavy $\Phi_{1,3}$ assumptions,  
the heavy scalars do not contribute to  the anomalous magnetic moment in a relevant way.

Moreover, we comment that the way of generating dimension 6 operators in the EFT model is not restricted to those
ones depicted in  Fig.~\ref{fig:EFTgeneration} (b) and (c). Since $\Phi_{2,3}$ have the same quantum number, the
scalar potential $V(\Phi_2, \Phi_3)_{\rm 2HDM} $ interactions are only weakly constrained and could induce large 
$h_2$-$h_3$ mixing. As  discussed 
in Appendix \ref{sec:CPoddUV}, and emphasized before, in the presence of large  $\lambda_6$ or $\mu_8$ terms, these mixing effects can lead to large 
contributions to the dimension 6 operator in the  EFT model. 
Let us stress, however,  that large $\lambda_6$ and $\mu_8$ terms can also induce 
large mixing between $h_2$-$\phi_R$. 
Such possibility beyond the scope of the EFT model and is in tension with our initial assumption that mixing between 
different scalars are all small. 

\subsection{Scalar interactions and relevant phenomenology}
The next constraint is the decay channels modified by scalar interactions. In the EFT model, $\phi_I$ decays
to $e^+ e^-$, while $\phi_R$ decays to $e^+ e^-$ with same coupling as $\phi_I$. $\phi_R$ can also 
decay to $\mu^+ \mu^-$ if kinematics allowed. With the scalar potential from UV model, e.g. 3HDM plus singlet
scalar in Appendix \ref{sec:CPoddUV}, there are a few phenomenologically relevant decay channels,  $\phi_R^{'} \to 
\phi_I^{'} \phi_I^{'}$, $H_2^0 \to  \phi_I^{'} \phi_I^{'}$ and $H_2^0 \to  \phi_R^{'} \phi_R^{'}$, where $\phi_{I,R}^{'}$ and
$H_2^0$ are mass eigenstates of CP-even (CP-odd) light scalars and SM Higgs.
According to the mixing matrix for
both CP-even (odd) scalars in Appendix \ref{sec:CPoddUV}, the triple scalar couplings between the mass eigenstates $\phi_{I,R}^{'}$ 
and $H_2^0$ can be calculated, 
 \begin{align}
\mathcal {L}_{\rm tri}= \left(\lambda_\phi -\lambda_6\left(\frac{\lambda_6}{4\lambda_2} + \frac{\mu_8 v_1}{\sqrt{2}\lambda_2 v_2 v_\phi}\right)\right)v_\phi \phi^{'2}_I\phi^{'}_R+ \left(\frac{\lambda_6}{2} +\frac{\mu_8 v_1}{\sqrt{2} v_2 v_\phi}\right)v_2 \left(\phi^{'2}_I H_2^0  +  \phi^{'2}_R H_2^0\right) .
\end{align}

First, from our benchmark, we have $v_\phi = 4.7$ GeV, $m_{\phi_I^{'}} \approx 15 $ MeV and $v_1 \ll v_\phi$.  
The CP-even scalar $\phi_R^{'}$ has a coupling which is about $10^{-4}  (10^{-3})$ to electrons (muons) respectively,
while its coupling to pairs of $\phi_I^{'}$ is $2\lambda_{\phi} v_\phi$. Thus,  $\phi_R^{'}$ will dominantly decay into 
$\phi_{I}^{'}$ pairs. Assuming $m_{\phi_R^{'}} \sim \sqrt{\lambda_{\phi}} v_\phi$,
the branching ratio of its decay into $e^- e^+$ ($\mu^+ \mu^-$) will be about $\sim 10^{-8}$
($10^{-6}$) respectively.
Then, the previous constraints on $\phi_R^{'} $ shown in Fig.~\ref{fig:independentcoupling} (b), which are based on the assumption $BR(\phi_R^{'} \to \ell^+ \ell^-)\sim 1$, 
should be revised. At low energy electron colliders, the relevant search channels are $e^+ e^- \to \gamma \phi'_R$ and
$e^+ e^- \to \phi_R^{'*} \to \phi'_I \phi'_I$,  governed by the electron coupling and $e^+ e^- \to \mu^+ \mu^- \phi'_R$ governed by the muon coupling. 
Since $\phi'_R \to \phi'_I \phi'_I \to 4e$, there are
multiple leptons in the final state. Although the BaBar experiment has searched for new physics in similar channels, 
for instance $e^+ e^- \to h' A'$, $h' \to A' A'$ and $A' \to \ell^+ \ell^-$~\cite{Lees:2012ra} and $e^+ e^- \to W' W' \to 2(\ell^+ \ell^-)$ 
in exclusive mode~\cite{Aubert:2009af}, it has not
explored the invariant mass regions consistent with $m_{\phi_I^{'}}$.  However, BaBar has the capability of lowering the
invariant mass threshold, as has been shown in the 2014 search for dark photons via $\gamma A'$ channel~\cite{Lees:2014xha}, 
where the BaBar collaboration extended 
the di-electron resonance channel to $m_{e^+ e^-} > 0.015$ GeV, and fits for $m_{A'} > 0.02$ GeV. We believe it would be
important to reanalyze their searches by imposing similar bounds on the dielectron invariant mass. 
Moreover, since $\phi_I^{'}$ and $ \phi_R^{'}$ are pretty light, they will be very boosted at high energy  colliders 
and form lepton jets \cite{ArkaniHamed:2008qp,Baumgart:2009tn,Bai:2009it,Katz:2009qq}. The proper lifetime of
$\phi_I^{'}$ in the benchmark is about $c\tau \approx  10^{-3} {\rm cm}$, thus it will appear as a prompt lepton jet in 
a low energy lepton collider, but displaced lepton jet at the  LHC. The displacement could help the search at the LHC, to separate the signal from the SM background, for example photon conversions. However, the invariant mass of the di-electron or even four lepton events coming from $\phi_R^{'}$ might be too low for the LHC experiments to 
detect them.

Second, we discuss the exotic SM Higgs decay channels $H_2^0 \to  \phi_I^{'} \phi_I^{'} \to 2(e^+e^-)$ and
$H_2^0 \to \phi_R^{'} \phi_R^{'} \to 4(e^+e^-)$. 
It is clear that if $\lambda_6$ is of $\mathcal{O}(1)$, then the SM-like Higgs will dominantly decay to those light scalars
thus one needs $\lambda_6 \ll 1$.  The ratio $\mu_8 v_1/(\sqrt{2} v_2 v_\phi)$ should also be small. To obtain a
$H_2^0 \to  \phi_I^{'} \phi_I^{'} , \phi_R^{'} \phi_R^{'}$ branching ratio smaller than $1\%$, the coefficient
$\lambda_6$ or $\mu_8 v_1/(\sqrt{2} v_2 v_\phi)$ should be $\lesssim 10^{-3}$, thus 
$\mu_8 v_1 \lesssim 1.7 ~{\rm GeV}^{2}$.  
If we tune down both $\mu_8$ and $v_1$, the $A_1^0 / H_1^0$ masses, of about $\sqrt{v_\phi v_2 (\mu_8/v_1)}$
(see Appendix \ref{sec:CPoddUV}), can still remain as heavy as $\sim 300$ GeV, with $\mu_8 \sim 10$ GeV and 
$v_1 \sim 0.1$ GeV. 
Interestingly, in the electron sector, we have $m_e = y_e v_1/\sqrt{2}$, which suggests 
$v_1 \gtrsim m_e$ and one can further decrease $v_1$ to make $A_1^0 / H_1^0$ heavier.  
Furthermore,  according to Eq.~(\ref{eq:gphiIeUV}), the coupling  $g_{\phi_I}^e$ is not affected by a small $v_1$.
One should note that, as we mentioned before, in the case $g_{\phi_R}^\mu$ is generated from triplet scalar interactions, 
we have from Eq.~(\ref{eq:gEFTmuphiR}) that for the benchmark presented in Table~\ref{tab:benchmark1},
$m_{h_3} = 37.8 \sqrt{y_{\mu 3} \mu_9 v_1} $. Hence,
if we take  $v_1 = 0.1$ GeV while keeping $\mu_9 \sim 1$ TeV and $y_{\mu 3}  \sim 1$, the mass $m_{h_3} $
goes down to $\sim 380$ GeV and will become smaller for smaller values of $v_1$. However, for the case $g_{\phi_R}^\mu$ is 
generated from quartic scalar interaction, Eqs. (\ref{eq:gEFTmuphiR4point}),
 the mass $m_{h_3}$ does not have a strong dependence on $v_1$ and hence
could remain heavy even for  very small values of $v_1$.

\subsection{Charged lepton flavor violation}
In this section, we discuss the possible flavor changing neutral current (FCNC) constraint.  Since the muon and the tau leptons
have the same quantum number, in the EFT Lagrangian, Eq.~(\ref{eq:EFT-Lag}), the muon
leptons can be substituted with tau leptons. Moreover, in the UV model, the two Higgs doublets $\Phi_{2,3}$
have the same quantum charge and hence admit the same couplings.  After the charged lepton mass
matrix diagonalization, a possible misalignment between the lepton mass and Yukawa couplings can induce
off-diagonal Yukawa couplings to muons and taus, see also a recent review \cite{Lindner:2016bgg} on $\Delta a_\mu$ and
lepton flavor violation.  
To avoid the appearance of FCNC, one can assume minimal flavor violation
(MFV) \cite{Chivukula:1987py, DAmbrosio:2002vsn} to align the couplings of $\Phi_{3}$ with the $\Phi_2$ ones. 
In the case of MFV, $\Phi_3$ will also couple to muon and tau lepton with diagonal couplings 
weighted by the lepton masses. Heavy Higgs bosons, which couple only to leptons and gauge bosons are difficult to test at 
hadron colliders. Under the MFV assumption,  however, the light scalar $\phi_R^{'}$ couples in a relevant 
way to $\tau$ leptons and is constrained to have a mass
between $30 - 200$ MeV in order to be consistent with precision electroweak 
constraints associated with  loop corrections to $Z \to \tau^+ \tau^-$~\cite{Abu-Ajamieh:2018ciu}. 

While MFV can solve the FCNC constraint for heavy scalars, the constraints on the light scalar couplings remain severe.
This is represented by the LFV decay $\tau \to \mu \phi'_R \to \mu + 2(e^+ e^-)$. The total width of $\tau$
is very small, $2.27 \times 10^{-12} $ GeV  and the  current limit on the three lepton LFV decay is
${\rm BR}(\tau \to \mu e^+e^-) < 1.8 \times 10^{-8}$~\cite{Tanabashi:2018oca}. This limit is easy to satisfy because 
$ {\rm BR}(\tau \to \mu e^+e^-) = {\rm BR}(\tau \to \mu \phi'_R)  \times {\rm BR}(\phi'_R \to  e^+e^-) $ for our 
benchmark point and ${\rm BR}(\phi'_R \to e^+e^-)\sim 10^{-8}$ as discussed above. 
However, since $\phi_R'$ can decay into pairs of  $\phi_I'$, there is a potential flavor violation in the channel 
$\tau \to  \mu + 2(e^+ e^-)$. We did not find limits on this channel  at the PDG~\cite{Tanabashi:2018oca}, but if 
the limits were of the same order as the one on ${\rm BR}(\tau \to \mu e^+e^-) $, it will imply 
$y_{\tau \mu }^{\phi_R} \lesssim 10^{-10}$. Since $y_{\tau \mu }^{\phi_R} = \sin \theta_{\phi_R h_3} y_{\tau \mu }^{h_3}$, 
and the mixing angle is about $10^{-3}$, one should restrict the LFV coupling $y_{\tau \mu }^{h_3} $ 
down to $10^{-7}$. Therefore, the alignment of the lepton Yukawa couplings must be enforced by a symmetry.
The most natural candidate would be  an extra  global $U(1)_\mu \times U(1)_\tau$ symmetry,
which is vector-like when applied to fermions unlike the chiral $U(1)_{\rm PQ}^e$. These symmetries forbid the off-diagonal 
terms between charged lepton species, and then the charged lepton mass matrix is  diagonal and LFV is not present
in the charged lepton sector.

\subsection{Others constraints and discussion}

Besides the FCNC issue, the Pontecorvo–Maki–Nakagawa–Sakata (PMNS) 
matrix for the lepton sector needs to be generated. Given the global 
$U(1)_{\rm PQ}^e \times U(1)_\mu \times U(1)_\tau$ symmetry, the Yukawa matrices 
of the SM charged and neutral leptons are diagonal. However, assuming a see-saw mechanism, 
one can generate the PMNS matrix from mixing  
in the heavy sterile neutrino sector \cite{Pascoli:2006ci,King:2013eh, Akhmedov:2013hec}, by assuming that the mass terms of the sterile neutrino 
$m^N_{ij} N^c_i N_j$ softly break the global symmetry  (see, for instance, the review, Ref.~\cite{Xing:2015fdg}, 
for the case of $U(1)_{\mu - \tau}$). 

Finally, we briefly mention that a $\phi'_I$ mass around 15 MeV, as required   to satisfy $\Delta a_e$ and the
other relevant phenomenological  constraints, is accidentally within the mass region necessary to 
explain the so-called  \ce{^{8}Be^{*}} anomaly, observed 
by the Atomki collaboration~\cite{Krasznahorkay:2015iga}.  Addressing this anomaly would imply a coupling of the
singlet scalar to quarks, something that is beyond the scope of our work. Let us stress, however, that 
the authors of Ref.~\cite{Ellwanger:2016wfe, Alves:2017avw} concluded that this possibility 
is subject to relevant constraints from low energy meson experiments that can only be avoided by assuming specific
coupling structures in the quark sector.

\section{Conclusions}
\label{sec:conclusions}

We have presented a scenario with a light complex scalar which can simultaneously accommodate
the anomalies in the electron and muon anomalous magnetic moments. The interesting feature is
that the same complex scalar induces positive contributions to $a_\mu$
and negative contributions to $a_e$.  This is achieved by assuming that the CP-even
component is much heavier than the CP-odd component and
having the CP-odd scalar scalar coupled only to electrons, while the CP-even couples to both
the electron and muon fields. This scenario may be realized in a natural way by introducing an
approximate PQ-like symmetry and assuming that the CP-odd scalar is a pseudo-Goldstone
boson associated to its spontaneous breakdown.  The EFT model
can then be written down directly and cope with the anomalies, while evading all the existing constraints.

We also analyzed how to generate such EFT model from a Standard Model extension containing 
multiple Higgs doublets. While the additional heavy Higgs doublet masses may be as large as 1~TeV,
flavor changing neutral currents may be avoided by assuming a global symmetry in the lepton
sector, broken softly in the neutrino sector. Furthermore, the
heavy scalars contribution to the 
anomalous magnetic moments is much smaller than the one of the light scalars
due to the small masses of the CP-odd and even component of
the complex scalar compared to the ones of the heavy Higgs bosons. For the light complex scalar,
its CP-odd and even components could be potentially reached by future B-factories and the HL-LHC. 
Looking for multiple prompt lepton jets in low energy electron collider and displaced lepton jets
from exotic SM Higgs decay at LHC is also a promising way to find those light scalars.

\section*{Acknowledgments}
Work at University of Chicago is supported in part by U.S. Department of Energy grant number DE-FG02-13ER41958. Work at ANL is supported in part by the U.S. Department of Energy under Contract No. DE-AC02-06CH11357.  The work  of CW was partially performed at the Aspen Center for Physics, which is supported by National Science Foundation grant PHY-1607611. We would like to thank Zhen Liu, Ian Low, Joshua T. Ruderman, Emmanuel Stamou, Lian-Tao Wang, and Neal Weiner for useful discussions and  comments. 
JL acknowledges support by Oehme Fellowship.

\newpage
\appendix
\section{The CP-even and CP-odd scalars in full UV model}
\label{sec:CPoddUV}
We consider the full UV model with three Higgs doublet $\Phi_{1,2,3}$ and one singlet complex scalar $\phi$,
where $\Phi_1$ and $\phi$ carries global $U(1)_{PQ}^{e}$ charge. The general scalar 
potential is
\begin{align}
V=&\mu_1^2 \Phi_1^\dag \Phi_1 + \mu_2^2 \Phi_2^\dag \Phi_2 + \mu_3^2 \Phi_3^\dag \Phi_3 + \mu_\phi^2 \phi^* \phi  
- m_{23}^2 \left( \Phi_2^\dag \Phi_3 + \Phi_3^\dag \Phi_2\right) + \frac{1}{2}\mu_4^2 \phi_I^2 \nonumber \\
& + \lambda_1 \left(\Phi_1^\dag \Phi_1\right)^2 + \lambda_2 \left(\Phi_2^\dag \Phi_2 \right)^2 + \lambda_{3} \left( \Phi_3^\dag \Phi_3\right)^2+ \lambda_\phi \left(\phi^* \phi \right)^2 + \lambda_4 \left( \Phi_1^\dag \Phi_1\right)\left( \Phi_2^\dag \Phi_2\right) \nonumber \\
&+ \lambda_5 \left( \Phi_1^\dag \Phi_1\right)\left( \phi^*\phi \right) + \lambda_6 \left( \Phi_2^\dag \Phi_2\right) \left( \phi^* \phi\right) + \lambda_7 \left(\Phi_2^\dag \Phi_1 \right)  \left(\Phi_1^\dag \Phi_2\right) + \lambda_8 \left( \Phi_3^\dag \Phi_3 \right) \left( \phi^* \phi\right) \nonumber \\
&+ \lambda_9 \left( \Phi_2^\dag \Phi_3  + \Phi_3^\dag \Phi_2\right) \left( \phi^* \phi\right)  +\mu_8 \left( \Phi_1^\dag \Phi_2 \phi^*+ H.c.\right) + \mu_9 \left( \Phi_1^\dag \Phi_3 \phi^*  + H.c.\right)  \nonumber \\
& + \lambda_{23}^a \left(\Phi_2^\dag \Phi_2\right)\left( \Phi_3^\dag \Phi_3\right)  
+  \lambda_{23}^b \left(\Phi_2^\dag \Phi_3\right)\left( \Phi_3^\dag \Phi_2\right)  \nonumber \\
&+  \lambda_{23}^c \left[\left(\Phi_2^\dag \Phi_3\right)^2 + \left( \Phi_3^\dag \Phi_2\right) ^2 \right] + \left(\lambda_{23}^d \Phi_2^\dag \Phi_2 + \lambda_{23}^e \Phi_3^\dag \Phi_3\right)\left(\Phi_2^\dag \Phi_3 +\Phi_3^\dag \Phi_2\right)   + \ldots
\end{align}
where we only written the scalar potential contributions, Eq.~(\ref{eq:LagPhi1}) and Eq.~(\ref{eq:LagPhi3}), which are relevant 
to the computation of $\Delta a_{e,\mu}$. The ``$\ldots$" denotes the irrelevant terms like 
$(\Phi_1^\dagger \Phi_1)(\Phi_3^\dagger \Phi_3)$ etc, which are neglected to avoid a too cumbersome computation.  

Minimizing the scalar potential, one obtains the following relations
\begin{align}
\mu_1^2 =&-\left[ \lambda_1 v_1^2 + \frac{\left(\lambda_4 +\lambda_7\right)}{2} v_2^2 +\frac{ \lambda_5}{2} v_\phi^2 +\frac{v_\phi}{\sqrt{2}v_1}  \left(\mu_8 v_2+\mu_9 v_3\right)\right] , \nonumber \\
\mu_2^2=&-\left[ \lambda_2 v_2^2 + \frac{\lambda_{23}^a + \lambda_{23}^b + 2\lambda_{23}^c}{2}v_3^2  +\frac{\left(\lambda_4 +\lambda_7\right) v_1^2}{2} + \frac{\lambda_6 v_\phi^2}{2} \right. \nonumber \\
&\left. + \frac{v_3}{2v_2}\left(3\lambda_{23}^d v_2^2 +\lambda_{23}^e v_3^2  + \lambda_9 v_\phi^2 - 2 m_{23}^2 \right) + \frac{\mu_8 v_1v_\phi}{\sqrt{2}v_2}\right]  ,\nonumber \\
\mu_3^2 =&-\left[ \lambda_3 v_3^2 + \frac{\lambda_{23}^a + \lambda_{23}^b + 2\lambda_{23}^c}{2}v_2^2  + \frac{v_2}{2v_3} \left(\lambda_{23}^d v_2^2 + 3\lambda_{23}^e v_3^2 - 2m_{23}^2\right)+\frac{v^2_\phi}{2v_3}\left(\lambda_8 v_3+\lambda_9 v_2  \right) + \frac{\mu_9 v_1 v_\phi}{\sqrt{2} v_3}\right] ,\nonumber \\
\mu_\phi^2  =& -\left[ \lambda_\phi v_\phi^2 +\frac{\lambda_5 v_1^2 +\lambda_6 v_2^2 +\lambda_8 v_3^2 + 2\lambda_9 v_2v_3 }{2} + \frac{v_1}{\sqrt{2} v_\phi }\left(\mu_8  v_2 + \mu_9  v_3\right) \right] .
\end{align}
We can diagonalize the mass matrix of CP-even or CP-odd scalars and obtain the mass in the leading order
under the assumption $v_3 \ll v_1 \ll v_\phi \ll v_2$ and $v_2 \sim \mu_{8,9} \sim m_{23}$.
The results for the CP-odd scalars are given by the eigenvalues
\begin{align}
m_{A^0_1}^2 & \approx -\frac{ v_\Phi}{\sqrt{2} v_1} \left(\mu_8 v_2 + \mu_9 v_3\right) , \\
m_{A^0_3}^2 & \approx \frac{ v_2\left(2 m_{23}^2 - \lambda_{23}^d v_2^2-\lambda_9 v_\phi^2\right) - \sqrt{2} \mu_9 v_1 v_\phi}{2 v_3} 
- 2 \lambda_{23}^c v_2^2  ,
\\
m_{\phi'_I}^2& \approx \mu_4^2 ,
\end{align} 
where $A^0_2$ is the massless Goldstone associated with the breakdown of the electroweak symmetry. 
$A_{1,2,3}^0$ and $\phi_{I}^{'}$ are the mass
eigenstates, while $a_{1,2,3}^0$ and $\phi_{I}$ are flavor states. 
If the results contain not only the leading terms, we always put the leading term on the left
and the sub-leading term on the right. The $4\times 4$ mixing matrix 
for CP-odd scalars in the leading order is given by,
\begin{align}
\left(\begin{array}{c} a_1 \\ a_2 \\ a_3 \\ \phi_I\end{array}\right) \approx
\left(\begin{array}{cccc}
1 &  \frac{v_1}{v} &  U_{13}^A & \frac{- v_1}{v_\phi} \\
- \frac{v_1}{v} & 1& -\frac{v_3}{v_2} & \frac{v_1^2}{v_2 v_\phi}\\
- U_{13}^A  & \frac{v_3}{v}  & 1 
& \mathcal{O}\left(\frac{v_{1,3, \phi}^3}{v_2^3} \right) \\
\frac{v_1}{v_\phi} & \mathcal{O}\left(\frac{v_{1,3, \phi}^3}{v_2^3} \right) & 
U_{43}^A & 1
\end{array}\right)
\left(\begin{array}{c} A^0_1 \\ G^0 \\ A^0_3 \\ \phi'_I\end{array}\right) ,
\end{align}
where $v=\sqrt{v_1^2 + v_2^2 + v_3^2}$, $v_1\ll v_\phi$, and $\mathcal{O}\left({v_{1,3, \phi}^3}/{v_2^3} \right)$ means
at least three orders in small parameter expansion.
\begin{align}
U_{13}^A &=\frac{\sqrt{2}\mu_9  v_\phi v_1}{2 m_{23}^2 v_1  - \lambda_{23}^d v_2^2v_1-v_\phi\left( \lambda_9v_1v_\phi -\sqrt{2}\mu_8 v_3\right)}   \frac{v_3}{v_2}  \simeq \frac{\sqrt{2}\mu_9  v_\phi}{2 m_{23}^2  - \lambda_{23}^d v_2^2}   \frac{v_3}{v_2}  ,\nonumber \\
U_{43}^A& =\frac{\sqrt{2} \mu_9 v_1^2 }{2 m_{23}^2 v_1 - \lambda_{23}^d v_2^2v_1 -\lambda_9 v_\phi^2 v_1+\sqrt{2}\mu_8 v_\phi  v_3}  \frac{v_3}{v_2}\simeq \frac{\sqrt{2} \mu_9 v_1  }{2 m_{23}^2 - \lambda_{23}^d v_2  } \frac{v_3}{v_2} \simeq \frac{v_1}{v_\phi}U_{13}^A.
\end{align}

The calculation for CP-even scalars are similar, with $H_{1,2,3}^0$ and $\phi_{R}^{'}$ being the mass
eigenstates, while $h_{1,2,3}^0$ and $\phi_{R}$ being the flavor states. The eigenvalues for CP-even
scalars are,
\begin{align}
m_{H^0_1}^2  \approx &- \frac{\left(v_\phi ^2 + v_1^2\right)}{\sqrt{2} v_1v_\phi}\left(\mu_8 v_2+\mu_9 v_3 \right) -\frac{\mu_8 v_\phi v_1}{\sqrt{2} v_2} \simeq - \frac{v_\phi}{\sqrt{2} v_1}\left(\mu_8 v_2+\mu_9 v_3 \right), \\
m_{H^0_2}^2  \approx &2 \lambda_2 v_2^2 
+ \frac{2 \lambda_{23}^d v_2 v_3 \left(3 \lambda_{23}^d v_2^2 -4 m_{23}^2 \right)}{\lambda_{23} v_2^2 - 2 m_{23}^2} , \\
m_{H^0_3}^2  \approx & \frac{\left(2 m_{23}^2  - \lambda_{23}^d v_2^2-\lambda_9 v_\phi^2\right)v_2}{2 v_3} -\frac{\mu_9 v_1 v_\phi}{\sqrt{2} v_3}+\frac{v_3 \left(2m_{23}^2-3\lambda_{23}^d v_2^2\right)^2}{2v_2\left(m_{23}^2  - \lambda_{23}^d v_2^2\right)}
 +\frac{3\lambda_{23}^e v_2v_3}{2}  \nonumber \\
 \approx  & \frac{2 m_{23}^2 v_2 - \lambda_{23}^d v_2^3}{2 v_3} + \mathcal{O}\left(\frac{v_{1,3, \phi}}{v_2} \right), \\
m_{\phi'_R}^2 \approx & \left(2 \lambda_\phi -\frac{\lambda_6^2}{2 \lambda_2} \right) v_\phi^2 
- \frac{\sqrt{2} \lambda_6 \mu_8 v_\phi v_1 }{\lambda_2 v_2} - \frac{\mu_8^2 v_1^2}{\lambda_2 v_2^2}.
\end{align} 
We see that under the hierarchical vevs assumption, $m_{A_1^0} \approx m_{H_1^0}$
and $m_{A_2^0} \approx m_{H_2^0}$, while $m_{\phi_R^{'}} > m_{\phi_I^{'}} $.
The mixing matrix for CP-even scalars is given by, 
\begin{align}
 \left(\begin{array}{c} h_1 \\ h_2 \\ h_3 \\ \phi_R \end{array}\right) \approx 
 \left(\begin{array}{cccc}
1 &  \frac{(2 \lambda_2 + \lambda_6) v_1}{2 \lambda_2 v_2} 
&  \frac{\sqrt{2} \mu_9   v_\phi}{2 m_{23}^2 - \lambda_{23}^d v_2^2}   \frac{v_3}{v_2} & \frac{v_1}{v_\phi} \\
- \frac{v_1}{v_2} & 1 & \frac{2 m_{23}^2 - 3 \lambda_{23}^d v_2^2}{\lambda_{23}^d v_2^2 -  2 m_{23}^2} \frac{v_3}{v_2}
 &- \frac{\sqrt{2}\mu_8  v_1+\lambda_6 v_\phi v_2}{2 \lambda_2 v_2^2} \\
\frac{\sqrt{2}\mu_9  v_\phi }{ \lambda_{23}^d v_2^2 - 2 m_{23}^2   }   \frac{v_3}{v_2} 
 &  \frac{ 3 \lambda_{23}^d v_2^2 - 2 m_{23}^2  }{\lambda_{23}^d v_2^2 -  2 m_{23}^2} \frac{v_3}{v_2}& 1 
 & U_{34} \\
- \frac{v_1}{v_\phi} &  \frac{\sqrt{2}\mu_8  v_1 +\lambda_6 v_\phi v_2}{2\lambda_2 v_2^2} 
& \frac{\sqrt{2} \mu_9 v_1 + 2\lambda_9  v_\phi v_2 }{2 m_{23}^2 -\lambda_{23}^d v_2^2} \frac{v_3}{v_2} & 1
\end{array}\right)
\left(\begin{array}{c} H^0_1 \\ H^0_2 \\ H^0_3 \\ \phi'_R \end{array}\right) , 
\end{align}
where we have
\begin{align}
U_{34} \approx \frac{1}{m_{H_3}^2} \left( -\sqrt{2} \mu_9 v_1 -\lambda_9 v_2 v_\phi
-\frac{m_{23}^2 \mu_8 v_1}{\sqrt{2} \lambda_2 v_2^2} - \frac{\lambda_6 m_{23}^2 v_\phi}{2 \lambda_2 v_2}
+ \frac{3 \lambda_{23}^d \mu_8 v_1}{2 \sqrt{2} \lambda_2} 
+ \frac{3 \lambda_{23}^d \lambda_6  v_2 v_\phi}{4 \lambda_2} 
\right)  \label{eq:U34} .
\end{align}
We see clearly that  the above $\mu_9 v_1$ ($\lambda_9 v_2 v_\phi$) in $U_{34}$ terms match with 
$\sin\theta_{\phi_R h_3}$ in Eq.~(\ref{eq:MixphiRh3}) and Eq.~(\ref{eq:quarticMIX}) 
from the $2\times 2$ mass matrix calculation. 
The last four terms with $\lambda_2 v_2^2$ in the denominator show additional contributions
to the mixing, whose effects can be tuned
down by further assuming $\lambda_6, ~\lambda_{23}^d \ll 1$ and $m_{23} < v_2$, while still
keeping the scalars $\Phi_{1,3}$ heavy.

\bibliography{referencelist}

\bibliographystyle{JHEP}   

\end{document}